\documentclass[fleqn,10pt]{wlscirep}
\usepackage[utf8]{inputenc}
\usepackage[T1]{fontenc}
\usepackage{bm}
\usepackage{subfig}
\usepackage{float}
\usepackage{graphicx}

\DeclareMathOperator*{\argmax}{argmax}
\DeclareMathOperator*{\argmin}{argmin}

\newcommand{\R}{\mathbb{R}}
\newcommand{\Prob}{\mathbb{P}}

\title{Convolutional GRU Network for Seasonal Prediction of the El Ni\~no-Southern Oscillation}

\author[1,*]{Lingda Wang}
\author[2,*]{Savana Ammons}
\author[2,*]{Vera Mikyoung Hur}
\author[3,*]{Ryan L. Sriver}
\author[1,*]{Zhizhen Zhao}
\affil[1]{Department of Electrical and Computer Engineering, University of Illinois Urbana-Champaign, Urbana, IL, 61820, USA}
\affil[2]{Department of Mathematics, University of Illinois Urbana-Champaign, Urbana, IL, 61801, USA}
\affil[3]{Department of Atmospheric Sciences, University of Illinois Urbana-Champaign, Urbana, IL, 61820, USA}

\affil[*]{\{lingdaw2, savanaa2, verahur, rsriver, zhizhenz\}@illinois.edu}

\begin{abstract}
Predicting sea surface temperature (SST) within the El Ni\~no-Southern Oscillation (ENSO) region has been extensively studied due to its significant influence on global temperature and precipitation patterns. Statistical models such as linear inverse model (LIM), analog forecasting (AF), and recurrent neural network (RNN) have been widely used for ENSO prediction, offering flexibility and relatively low computational expense compared to large dynamic models. However, these models have limitations in capturing spatial patterns in SST variability or relying on linear dynamics. Here we present a modified Convolutional Gated Recurrent Unit (ConvGRU) network for the ENSO region spatio-temporal sequence prediction problem, along with the Ni\~no 3.4 index prediction as a down stream task. The proposed ConvGRU network, with an encoder-decoder sequence-to-sequence structure, takes historical SST maps of the Pacific region as input and generates future SST maps for subsequent months within the ENSO region. To evaluate the performance of the ConvGRU network, we trained and tested it using data from multiple large climate models, including a pre-industrial simulation spanning approximately 1300 years from the Community Climate System Model version 4 (CCSM4) and a 30-member historical hindcast ensemble during the years 1921-2100 using the NOAA Seamless system for Prediction and EArth System Research (SPEAR) model. We also compare and contrast the prediction skill of the ConvGRU network against existing models. The results demonstrate that the ConvGRU network significantly improves the predictability of the Ni\~no 3.4 index compared to LIM, AF, and RNN. This improvement is evidenced by extended useful prediction range, higher Pearson correlation, and lower root-mean-square error. The proposed model holds promise for improving our understanding and predicting capabilities of the ENSO phenomenon and can be broadly applicable to other weather and climate prediction scenarios with spatial patterns and teleconnections.
\end{abstract}
\begin{document}

\flushbottom
\maketitle
\thispagestyle{empty}

\section*{Introduction}
The El Ni\~no-Southern Oscillation (ENSO) phenomenon over the tropical Pacific region is the most energetic driver of climate variability on the seasonal to interannual timescales~\cite{mcphaden2006enso}. It plays a crucial role on global oceanic and atmospheric dynamics, particularly during its irregular warming (El Ni\~no) and cooling (La Ni\~na) phases. The impacts of ENSO are widespread, leading to anomalous temperature and precipitation patterns on a global scale~\cite{fraedrich1992climate, kumar1999weakening, glantz2001currents}, as well as changing extreme and hazardous weather conditions on a regional scale, such as winter to early spring tornado outbreaks in the United States~\cite{cook2017impact}, tropical cyclone intensity changes in northwestern Pacific~\cite{camargo2005western}, and unusual fire weather in Australia~\cite{squire2021likelihood} and the United States~\cite{trouet2009interannual}. Consequently, accurate prediction of sea surface temperature (SST) maps within the ENSO region and its associated Ni\~no indices---for instance, Niño 1+2, 3, 3.4, and 4~\cite{rasmusson1982variations,trenberth1997definition,trenberth2001indices}---has become a critical area of research. Reliable ENSO prediction can provide valuable insights for decision-making processes in various sectors, including government agencies, food and insurance industries, and transportation, enabling them to prepare for the associated impacts~\cite{hansen2006translating, hammer2001advances}.

Prediction models for SST maps within the ENSO region and the associated Ni\~no indices can be broadly classified into two types: dynamical models and statistical models. Dynamical models, such as the North American Multi-Model Ensemble (NMME)~\cite{barnston2019deterministic}, are commonly used for seasonal ENSO prediction. However, these model ensembles are computationally intensive, sensitive to initialization conditions, and expensive to run. In contrast, this study focuses on statistical models due to their simplicity and comparable prediction skill to dynamical models. One widely used statistical ENSO prediction model is the linear inverse model (LIM)~\cite{penland1993prediction, penland1995optimal}, which employs principal components analysis and Markov prediction to approximate trends and predict future states based on empirical orthogonal functions, similar to linear regression (LR)~\cite{wang2020extended}. However, LIM fails to capture nonlinear ENSO dynamics---for instance, surface-subsurface interactions and surface winds~\cite{chapman2015vector}---and can lead to underestimation of ENSO variability. Another type of statistical prediction model is based on Lorenz's analog forecasting (AF)~\cite{lorenz1969atmospheric}. Initially, AF based models use observed or free-running model data as libraries of states. Predictions are then generated by matching states in the library that are very similar to observed data at prediction initialization, and following the evolution on these so-called analogs. Advantages of AF based models include avoiding expensive and unstable initialization systems and reducing structural model error. The kernel analog forecasting (KAF) model~\cite{zhao2016analog,wang2020extended,burov2021kernel}, as a generalization of conventional AF based models, utilizes nonlinear kernels to better capture nonlinearity in ENSO dynamics. Recently, with the development of deep learning techniques, the convolutional neural network (CNN)~\cite{ham2019deep} and long short-term memory (LSTM)~\cite{hochreiter1997long, huang2019analyzing} network have been used for predicting Ni\~no indices, but their prediction skills have not yet been extended to capturing spatial patterns in SST variability within the ENSO region.

In this study, we propose the use of a Convolutional Gated Recurrent Unit (ConvGRU) network, inspired by and modified from the original developments~\cite{shi2015convolutional, shi2017deep, ballas2015delving}, to predict SST maps within the ENSO region, along with the Ni\~no 3.4 index as a downstream task. The ConvGRU network has an encoder-decoder sequence-to-sequence (Seq2Seq) structure, with both the encoder and the decoder consisting of multi-layer ConvGRU cells. The encoder compresses the input SST maps of the Pacific region into hidden states across all layers, and the decoder unfolds the hidden states from the encoder to generate predictions within the ENSO region. The ConvGRU cell, a key component of both the encoder and the decoder, incorporates several 2-D convolutional layers. This architecture enables the ConvGRU network to take historical SST maps of the Pacific region as inputs and generate future SST maps of the ENSO region for subsequent months, taking into consideration of spatio-temporal correlation of the SST maps. Moreover, this architecture significantly reduces the number of network parameters while accelerating the training process. 

To evaluate the performance of the ConvGRU network, we conduct numerical experiments and compare it against existing models, such as KAF, LIM, Seq2Seq with GRU, and LR, using global climate ensembles and atmospheric reanalysis datasets. These datasets include two SST datasets and one surface air temperature dataset. The comparison results demonstrate that the ConvGRU network achieves significant improvements over the other models in terms of useful prediction range, Pearson correlation (PC), and root-mean-square error (RMSE).

By developing an improved prediction model that accurately captures the complex dynamics and spatial patterns of SST within the ENSO region, this study aims to contribute to better understanding and prediction of ENSO-related climate phenomena. Further research can explore further enhancements to the network architecture and investigate its applicability to other climate-related features and prediction tasks.

The rest of this paper is organized as follows: In preliminaries, we provide an overview of the ENSO region prediction problem and discuss existing models for spatio-temporal sequence prediction. In methodology, we present the ConvGRU network, including the ConvGRU cells and the encoder-decoder Seq2Seq structure, and describe the training process. In results and discussion, we discuss the performance of the ConvGRU network on the ENSO region spatio-temporal sequence prediction task as well as its comparison with existing models.  

\section*{Preliminaries}

\subsection*{ENSO region prediction problem}

We address the ENSO region prediction problems, which involves predicting future SST map sequences within the ENSO region of the Pacific, given previously observed gridded SST maps of the Pacific region. Suppose that SST maps of the Pacific region are sampled and averaged monthly on a grid of size $M\times N$, representing an SST map of the Pacific region as a matrix in $\R^{M\times N}$ for a specific month. As monthly records of SST maps of the Pacific region are accumulated, a sequence of such matrices is obtained,  $\tilde{\bm{X}}_1,\ldots,\tilde{\bm{X}}_t,\ldots (\in \R^{M\times N})$, where $t$ denotes a specific month. Given the previous $J$-month (referred to as the condition range) observed SST maps of the Pacific region, including the current one, represented as $\tilde{\bm{X}}_{t-J+1:t}\in\R^{J\times M\times N}$, the ENSO region spatio-temporal sequence prediction problem at month $t$ aims to predict the most likely $K$-month (referred to as the prediction range) future SST maps within the ENSO region. These predicted maps are denoted as $\hat{\bm{Y}}_{t+1},\ldots, \hat{\bm{Y}}_{t+K} (\in \R^{M\times N})$, abbreviated as $\hat{\bm{Y}}_{t+1:t+K}\in\R^{K\times M\times N}$. Formally, the problem can be stated as follows:
\begin{equation}\label{eq:seq2seq}
    \hat{\bm{Y}}_{t+1:t+K} = \argmax_{\bm{Y}_{t+1:t+K}}\,\Prob\left(\bm{Y}_{t+1:t+K}\;\middle|\;\tilde{\bm{X}}_{t-J+1:t}\right).
\end{equation}
The extent of the ENSO region for the predicted maps can cover the entire Pacific region or any other region within the Pacific, depending on the downstream task, for example, the south Pacific decadal oscillation~\cite{zhang1997enso}. In real-world applications, SST maps of the Pacific region are typically sampled and averaged monthly on a latitude-longitude grid of a specific resolution, such as $1^\circ \times 1^\circ$ per latitude-longitude grid cell, and the prediction range spans 12 and 24 months. 

The ENSO region prediction problem encompasses a series of downstream tasks, such as predicting Ni\~no indices. It holds potential applications in other climate-related features such as fire weather and drought indices~\cite{hess2001nino, squire2021likelihood}.

\subsection*{Models for spatio-temporal sequence prediction}
There exists a range of approaches for spatio-temporal sequence prediction, including machine learning and traditional statistical models. We categorize these models into autoregressive models and statistical models, highlighting their applicability to weather and climate related tasks. 

\subsubsection*{Autoregressive models}

Autoregressive models have found widespread usage in time series prediction problems, including recurrent neural networks (RNNs) such as vanilla RNN~\cite{medsker2001recurrent}, LSTM~\cite{hochreiter1997long}, and GRU~\cite{cho2014properties}. One multivariate variant of general-purpose RNN, known as fully-connected RNN (FC-RNN)~\cite{graves2013generating, shi2015convolutional}, was among the earliest models employed for spatio-temporal sequence prediction. For instance, FC-LSTM takes vectorized inputs (spatio maps) and utilizes LSTM cells. The main equations for FC-LSTM can be summarized as follows:
\[
\begin{aligned}
&i_t = \sigma_i(W_{ix}x_t+W_{ih}h_{t-1}+b_i), &&\bullet\text{Input gate}\\
&f_t = \sigma_f(W_{fx}x_t+W_{fh}h_{t-1}+b_f),\qquad &&\bullet\text{Forget gate}\\
&o_t = \sigma_o(W_{ox}x_t+W_{oh}h_{t-1}+b_o), &&\bullet\text{Output gate}\\
&\tilde{c}_t = \sigma_{\tilde{c}}(W_{\tilde{c}x}x_t+W_{\tilde{c}h}h_{t-1}+b_{\tilde{c}}), &&\bullet\text{New memory cell}\\
&c_t = f_t \odot c_{t-1} + i_t \odot \tilde{c}_t, &&\bullet\text{Final memory cell}\\
&h_t = o_t \odot \sigma_h(c_t), &&\bullet\text{Hidden state}
\end{aligned}
\]
where $\odot$ represents the element-wise product (Hardmard product), and $\sigma$ is either the sigmoid or $\operatorname{tanh}$ function.

However, FC-LSTM has limitations in efficiently capturing spatial correlations. To overcome this drawback, ConvLSTM~\cite{shi2015convolutional} was introduced, which incorporates 2-D convolutional layers within an LSTM cell. ConvLSTM has been further enhanced with various variants and successors such as  TrajGRU~\cite{shi2017deep}, CDNA~\cite{finn2016unsupervised}, PredRNN~\cite{wang2022predrnn}. Another approach to autoregressive models for spatio-temporal sequence prediction is based on Transformers~\cite{fonseca2023continuous, yang2021tctn}. Additionally, autoregressive models have been combined with other techniques, such as graph neural networks~\cite{keisler2022forecasting, lam2022graphcast} and generative models~\cite{ravuri2021skilful}, leading to significant achievements in short to medium-range weather prediction and other spatio-temporal sequence prediction applications.

\subsubsection*{Statistical climate models}

Statistical climate models employ statistical techniques tailored for climate-related data analysis and prediction. Examples include LIM~\cite{penland1993prediction, penland1995optimal} and KAF~\cite{wang2020extended, zhao2016analog, burov2021kernel}. 

LIM assumes that the dynamics of a system can be described by a linear stochastic differential equation of the form:
\begin{equation*}
    \frac{d\bm{x}}{dt} = \bm{B}\bm{x}+\xi.
\end{equation*}
Here $\bm{x}(t)$ represents the state of the system at time $t$, $\bm{B}$ is a time-independent operator, and $\xi$ is stationary white noise. For stationary statistics, $\bm{B}$ must be dissipative, meaning that its eigenvalues have negative real parts, and 
\[
\bm{C}(\tau) = \bm{G}(\tau)\bm{C}(0)\quad\text{and}\quad \bm{G}(\tau)=\exp(\bm{B}\tau),
\]
where $\bm{C}(0)$ and $\bm{C}(\tau)$ are covariances of $\bm{x}$ at lags $0$ and $\tau$, respectively. In prediction problems, $\bm{G}(\tau)\bm{x}(t)$ represents the best linear prediction of the state at time $t+\tau$, given the state at time $t$. The matrices $\bm{B}$ and $\bm{G}$ can then be determined as $\bm{B} = \tau^{-1}\ln{(\bm{C}(\tau)C(0)^{-1})}$.

KAF is a generalization of AF~\cite{lorenz1969atmospheric, ding2018skillful}, incorporating both nonlinear kernel methods and operator-theoretic ergodic theory~\cite{eisner2015operator}. By establishing a rigorous connection with Koopman operator theory~\cite{giannakis2019data} for dynamical systems, KAF can generate statistically optimal predictions as conditional expectations. KAF is particularly useful when dealing with noisy and partially observed data during prediction initialization.

These models form the foundation for spatio-temporal sequence prediction and have been applied to various weather and climate problems. However, we should notice that i) LIM is not capable of capturing nonlinear dynamics with the ENSO region; ii) KAF is usually applied to Ni\~no index prediction task, and is not capable of predicting the spatial pattern within the ENSO region. Those limitations of statistical models motivate the proposal of the ConvGRU network in this study.

\section*{Methodology}

We now propose our ConvGRU network, which is inspired by and modified from the ConvGRU model~\cite{ballas2015delving, shi2017deep}, for the ENSO region spatio-temporal sequence prediction. The ConvGRU network incorporates 2-D convolutional layers in both input-to-(hidden) state and (hidden) state-to-(hidden) state transitions within a ConvGRU cell. This modification offers several advantages over FC-GRU cells, efficiently capturing spatial correlations of SST maps and reducing the number of network parameters. The ConvGRU network is composed of multiple stacked ConvGRU cells and follows an encoder-decoder Seq2Seq structure. During the training process, samples are generated from fixed-length windows with different starting points. 

\subsection*{Convolutional GRU cell}

\begin{figure}[t!]
    \begin{center}
        \includegraphics[width = 0.5\linewidth]{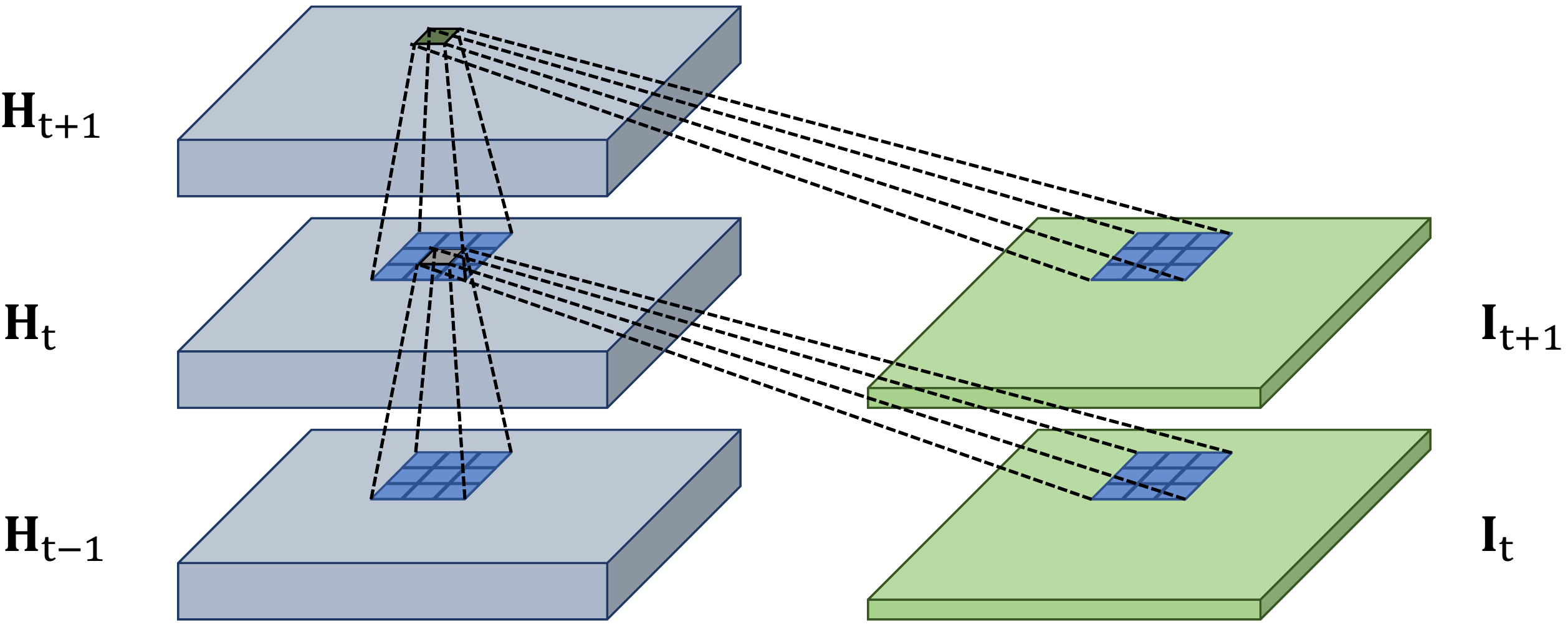}
    \end{center}
    \caption{Illustration of 2-D convolutional layers within a ConvGRU cell. Convolutional layers are applied to update gate, reset gate, and new memory cell (see Eq.~\eqref{eq:ConvGRU_cell}).}
    \label{fig:ConvGRU_cell}
\end{figure}

If we were to tackle the ENSO region spatio-temporal sequence prediction problem in Eq.~\eqref{eq:seq2seq} using a network built with FC-GRU cells, we would need to vectorize inputs and hidden states before performing matrix multiplication. These steps are essentially equivalent to fully connected layers in a neural network. However, the vectorization and matrix multiplication steps are not required when using ConvGRU cells. Instead, a ConvGRU cell employs 2-D convolutional layers, which offers several advantages, including extracting meaningful spatial correlation features, reducing the number of network parameters, and speeding up the training process.

Following the structure of FC-GRU cells, the equations for the ConvGRU cell can be expressed as follows:
\begin{equation}\label{eq:ConvGRU_cell}
\begin{aligned}
&z_t = \sigma_z(\bm{W}_{zx}\ast \bm{I}_t + \bm{W}_{zh}\ast \bm{H}_{t-1}+b_{z}), && \bullet\text{Update gate}\\
&r_t = \sigma_r(\bm{W}_{rx}\ast \bm{I}_t + \bm{W}_{rh}\ast \bm{H}_{t-1}+b_{r}), && \bullet\text{Reset gate}\\
&\tilde{\bm{H}}_t = \sigma_{\tilde{h}}(\bm{W}_{\tilde{h}x}\ast \bm{I}_t + \bm{W}_{\tilde{h}rh}\ast(r_t\odot \bm{H}_{t-1}) + b_{\tilde{h}}), \qquad && \bullet\text{New memory cell}\\
&\bm{H}_t = (1-z_t)\odot \tilde{\bm{H}}_t + z_t \odot \bm{H}_{t-1}, && \bullet\text{Hidden state}
\end{aligned}
\end{equation}
where $\ast$ represents the convolution operator. Here, input $\bm{I}_t$, hidden state $\bm{H}_t$, update gate $z_t$, reset gate $r_t$, and new memory cell $\tilde{\bm{H}}_t$ are all 3D tensors, with the last two dimensions representing the spatial dimensions (rows and columns). Figure~\ref{fig:ConvGRU_cell} illustrates the application of 2-D convolutional layers within a ConvGRU cell 
for both the input-to-(hidden) state and (hidden) state-to-(hidden) state transitions. This allows the future hidden state in a specific grid cell to extract relevant information locally from its neighboring inputs and past hidden states. The size of the neighbors considered by a grid cell is determined by the size of the convolutional kernel. A large kernel is recommended for fast-evolving spatio-temporal sequences, while a small kernel is more suitable for slow-varying sequences.

\subsection*{Encoder-Decoder Seq2Seq structure}

\begin{figure}[t!]
    \centering
    \subfloat[Encoder]{\includegraphics[width = 0.6\linewidth]{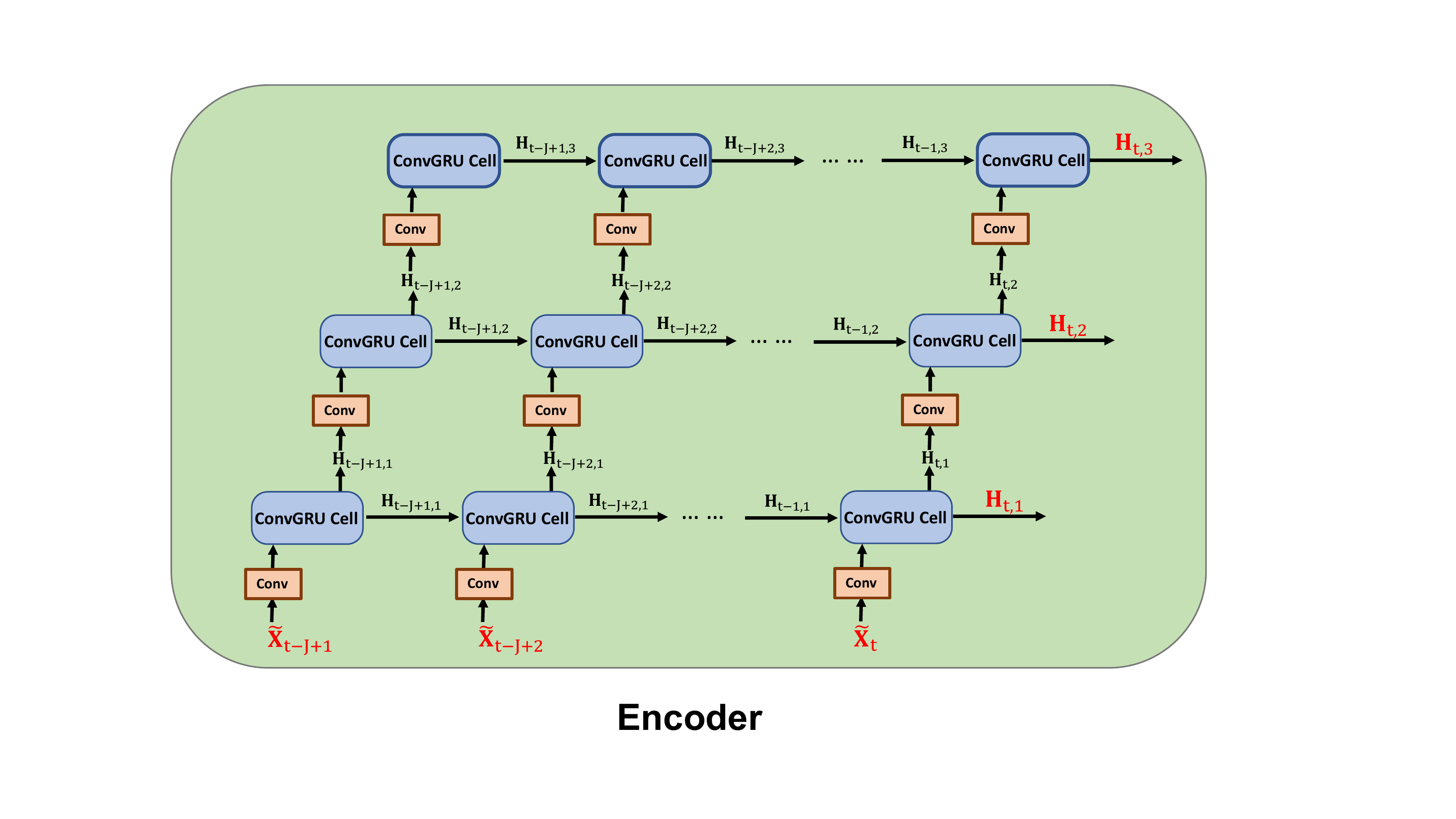}
    \label{fig:ENC}
    }\\
    \subfloat[Decoder]{\includegraphics[width = 0.6\linewidth]{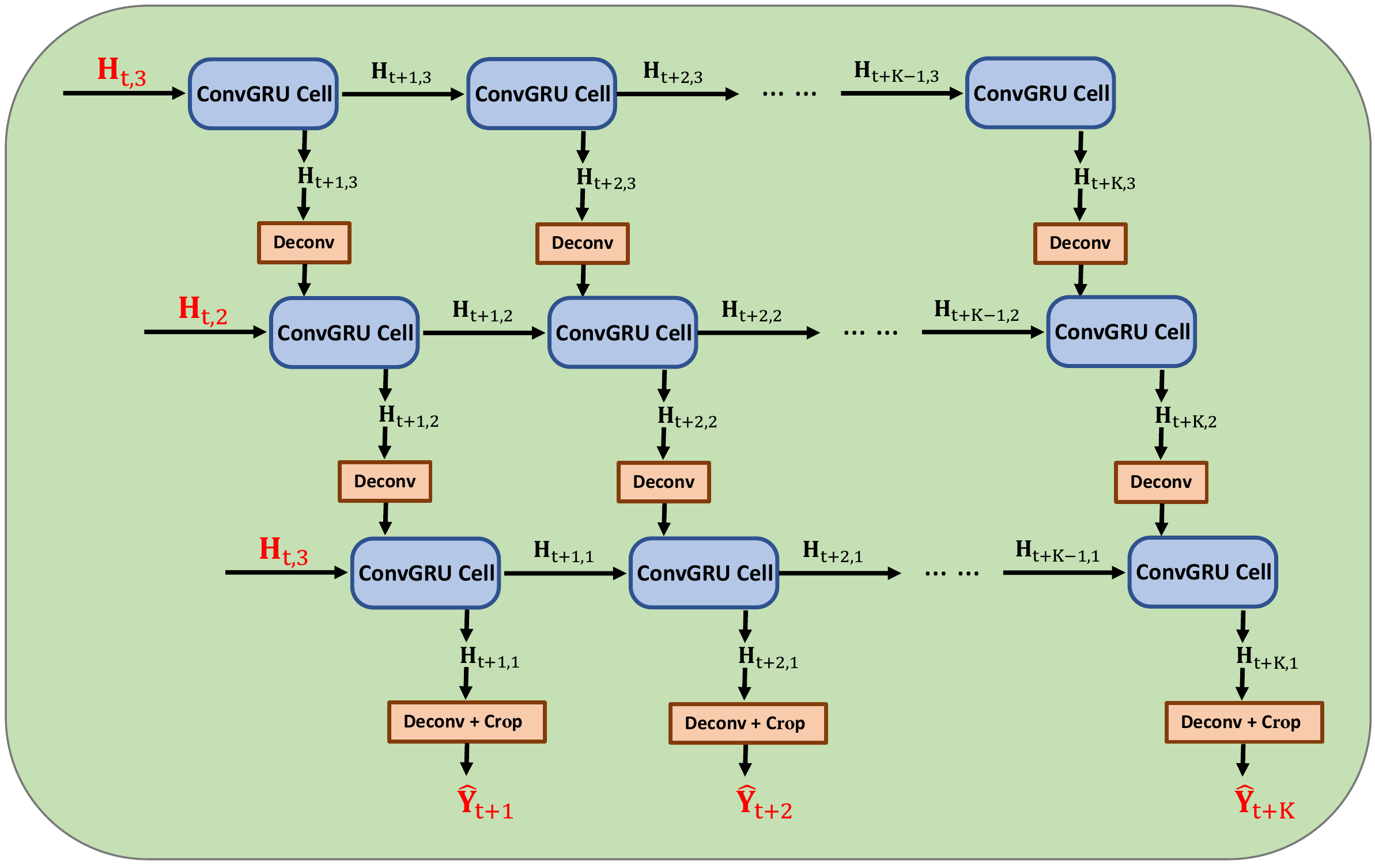}
    \label{fig:DEC}}
    \caption{Three-layer ConvGRU network, where the initial hidden states of the decoder are copied from the last hidden states of the encoder. \protect \subref{fig:ENC} Encoder architecture utilizing ConvGRU cells and 2-D convolutional layers. \protect \subref{fig:DEC} Decoder architecture constructed with ConvGRU cells and 2-D deconvolutional layer.}
    \label{fig:ED}
\end{figure}

We utilize the ConvGRU cells in Eq.~\eqref{eq:ConvGRU_cell} as a key component to construct our ConvGRU network for ENSO region spatio-temporal sequence prediction. We recognize this as a Seq2Seq learning problem (see Eq.~\eqref{eq:seq2seq}) that can be effectively addressed using the encoder-decoder Seq2Seq structure~\cite{cho2014properties,chung2014empirical}.

Figure~\ref{fig:ED} illustrates the architecture of the ConvGRU network for a 3-layer example, although the number of layers can be adjusted based on performance considerations, such as RMSE. The ConvGRU network consists of two main parts: multi-layer encoder and multi-layer decoder. 

The encoder, depicted in Figure~\ref{fig:ENC}, utilizes ConvGRU cells and 2-D convolutional layers. Each layer's hidden states are initialized as an all $0$-tensor and updated using inputs and the previous hidden states, following Eq.~\eqref{eq:ConvGRU_cell}. The first layer takes SST maps of the Pacific region as inputs, while the subsequent layers receive the output hidden states from the previous layer. Convolutional layers are applied before each ConvGRU cell, to adjust the number of the input channels and the size of the input spatial dimensions, enhancing feature extraction.

The decoder, depicted in Figure~\ref{fig:DEC}, consists of ConvGRU cells and 2-D deconvolutional layers. A crucial step that connects the encoder and decoder is that the hidden states of each layer in the decoder is copied from the last output hidden states of the corresponding layer in the encoder. The decoder architecture is similar to the encoder, but the flow direction of hidden states among layers is reversed. This enables the adoption of 2-D deconvolutional layers, which are the reverse operation of 2-D convolutional layer. They ensure that the inputs and network parameters in each decoder layer's ConvGRU cells are consistent with those in the encoder, so that the last output hidden states from the encoder can be utilized by the decoder. For the outputs of the first layer of the decoder, the grid is cropped to the ENSO region. 

The encoder-decoder Seq2Seq structure of the ConvGRU network can be interpreted as follows: the encoder compresses the input SST maps of the Pacific region into hidden states across all layers, while the decoder unfolds the hidden states from the encoder to generate predictions for the ENSO region. Consequently, the ConvGRU network approximates the problem stated in Eq.~\eqref{eq:seq2seq}:
\begin{equation}\label{eq:approximation}
\begin{aligned}
\hat{\bm{Y}}_{t+1:t+K} &= \argmax_{\bm{Y}_{t+1:t+K}}\,\Prob\left(\bm{Y}_{t+1:t+K}\;\middle|\;\tilde{\bm{X}}_{t-J+1:t}\right)\\
&\approx\argmax_{\bm{Y}_{t+1:t+K}}\, \Prob\left(\bm{Y}_{t+1:t+K}\;\middle|\;f_\text{ENC}\left(\tilde{\bm{X}}_{t-J+1:t}\;\middle|\;\bm{W}_\text{ENC}\right)\right)\\
&\approx g_\text{DEC}\left(f_\text{ENC}\left(\tilde{\bm{X}}_{t-J+1:t}\;\middle|\;\bm{W}_\text{ENC}\right)\;\middle|\;\bm{W}_\text{DEC}\right),
\end{aligned}
\end{equation}
where $f_\text{ENC}(\cdot|\bm{W}_\text{ENC})$ and $g_\text{DEC}(\cdot|\bm{W}_\text{DEC})$ represent the encoder and decoder, respectively, with network parameters $\bm{W}_\text{ENC}$ and $\bm{W}_\text{DEC}$.

\subsection*{Training process}

\begin{figure}[t!]
    \centering
    \subfloat[Training]{\includegraphics[width = 0.4\linewidth]{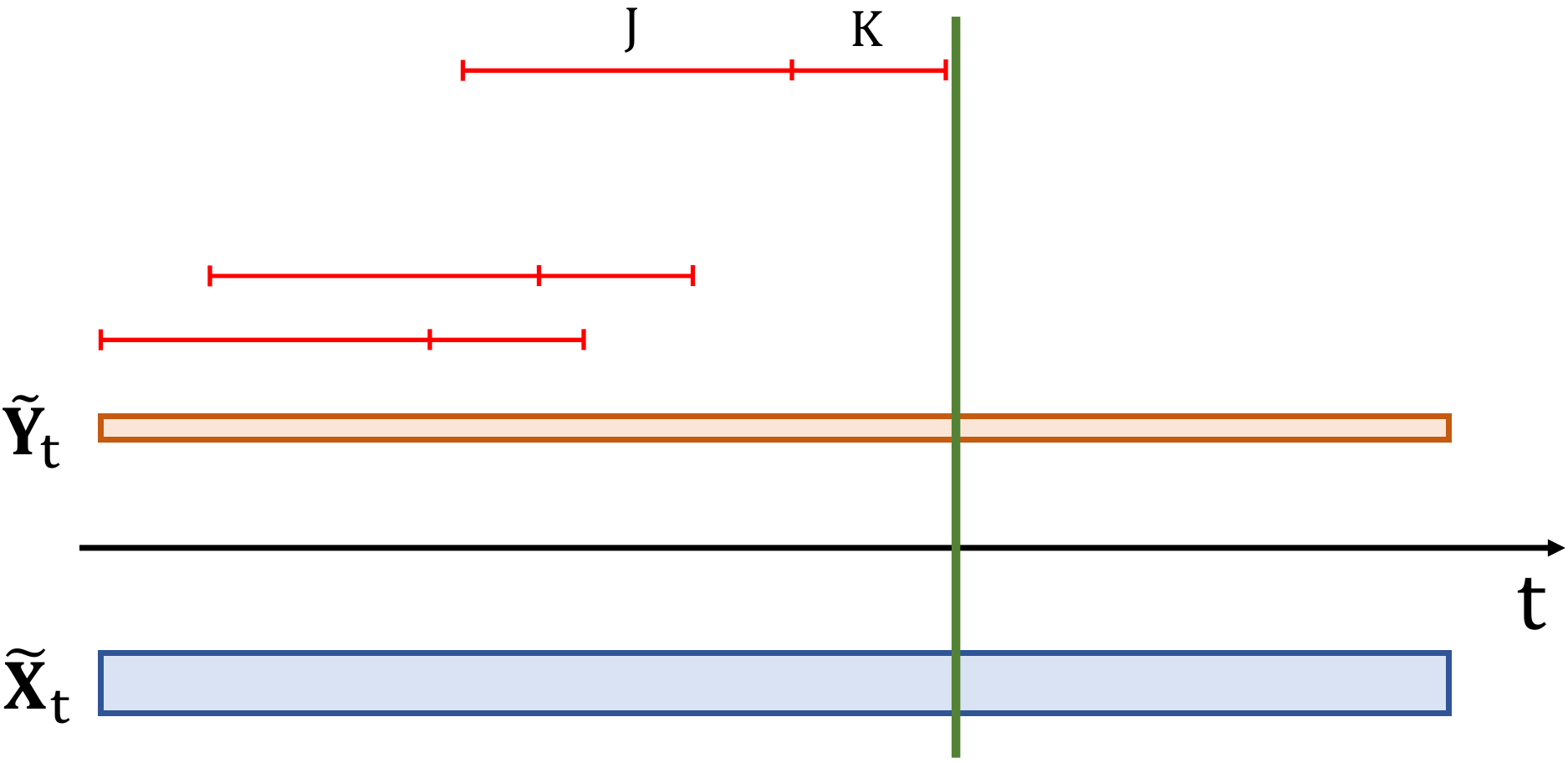}
    \label{fig:train}
    }\hspace{0.5cm}
    \subfloat[Testing]{\includegraphics[width = 0.41\linewidth]{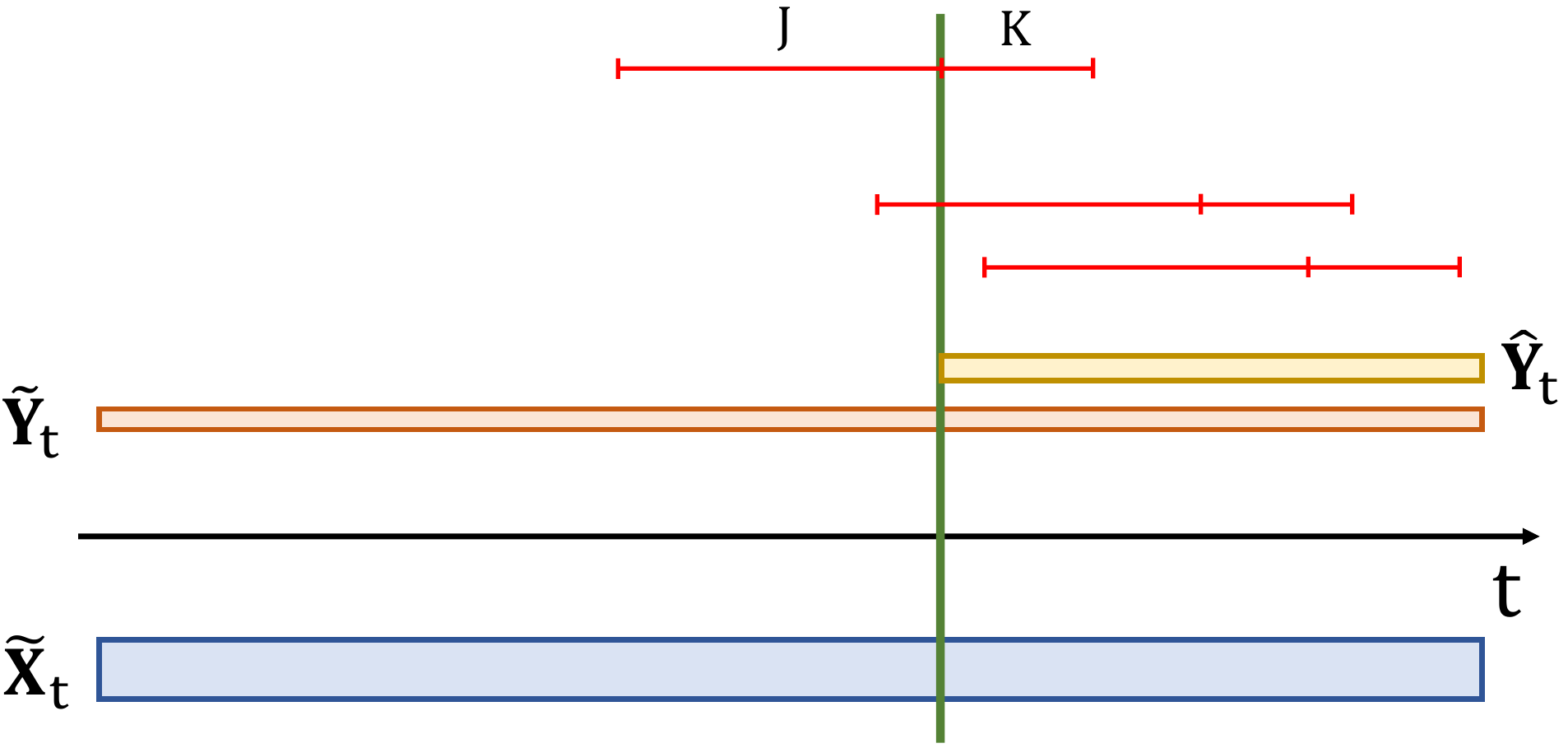}
    \label{fig:test}}
    \caption{Data setup for training and testing. The green vertical lines divide the entire dataset into the training data and the testing data. The performance metric is evaluated to the right of the green line, and no data in this part is used in training. \protect \subref{fig:train} Data setup during the training process. The red lines depict the training windows of $\{(\tilde{\bm{X}}_t, \tilde{\bm{Y}_t})\}_{t=1}^T$, where the left part represents the condition range ($t-J+1$ to $t$ for some $t$), and the right part represents the prediction range ($t+1$ to $t+K$). Note that all training windows are to the left of the green line. \protect \subref{fig:test} During the testing process, the prediction range is strictly to the right of the green line.}
    \label{fig:trainP}
\end{figure}

Given a training dataset consisting of SST maps, denoted as $\{(\tilde{\bm{X}_t}, \tilde{\bm{Y}_t})\}_{t=1}^T$, of the Pacific and ENSO regions, respectively, the network parameters $\bm{W}^*_\text{ENC}$ and $\bm{W}^*_\text{DEC}$ of the encoder and decoder can be learned by minimizing the difference between the predicted sequence $\hat{\bm{Y}}_{t+1:t+K}$ and the ground truth sequence $\tilde{\bm{Y}}_{t+1:t+K}$. The optimization process can be described as follows:
\begin{equation}
    \label{eq:cost_func}
    \begin{aligned}
    \bm{W}^*_\text{ENC}, \bm{W}^*_\text{DEC}&=\argmin_{\bm{W}_\text{ENC}, \bm{W}_\text{DEC}}\,\sum_{t=J}^{T-K}\mathcal{L}\left(\tilde{\bm{Y}}_{t+1:t+K},g_\text{DEC}\left(f_\text{ENC}\left(\tilde{\bm{X}}_{t-J+1:t}\;\middle|\;\bm{W}_\text{ENC}\right)\;\middle|\;\bm{W}_\text{DEC}\right)\right)\\
    &=\argmin_{\bm{W}_\text{ENC}, \bm{W}_\text{DEC}}\,\sum_{t=J}^{T-K}\mathcal{L}\left(\tilde{\bm{Y}}_{t+1:t+K}, \hat{\bm{Y}}_{t+1:t+K}\right),
    \end{aligned}
\end{equation}
where the loss function $\mathcal{L}$ can be chosen, for instance, as the mean squared loss. The optimization problem in Eq.~\eqref{eq:cost_func} can be solved using stochastic gradient descent algorithms, such as Adam~\cite{kingma2014adam} and Adagrad~\cite{duchi2011adaptive}.  

During the training process, multiple training windows (instances) of length $J+K$ are generated from the training dataset, each with different start points. The condition ($J$) and prediction ($K$) ranges remain fixed for all training windows. For instance, if the training dataset spans from month $1$ to month $10000$, training windows can be created with $t$ in Eq.~\eqref{eq:cost_func} ranging from $J$ to $10000-K$. Figure~\ref{fig:train} illustrates the generation of training windows. Once the ConvGRU network is trained, it can be evaluated on a testing dataset  $\{(\tilde{\bm{X}_t}, \tilde{\bm{Y}_t})\}_{t=T+1}^{T+F}$, as depicted in Figure~\ref{fig:test}.

\begin{figure}[htb!]
    \centering
    \subfloat[Ground truth]{\includegraphics[width = 0.45\linewidth]{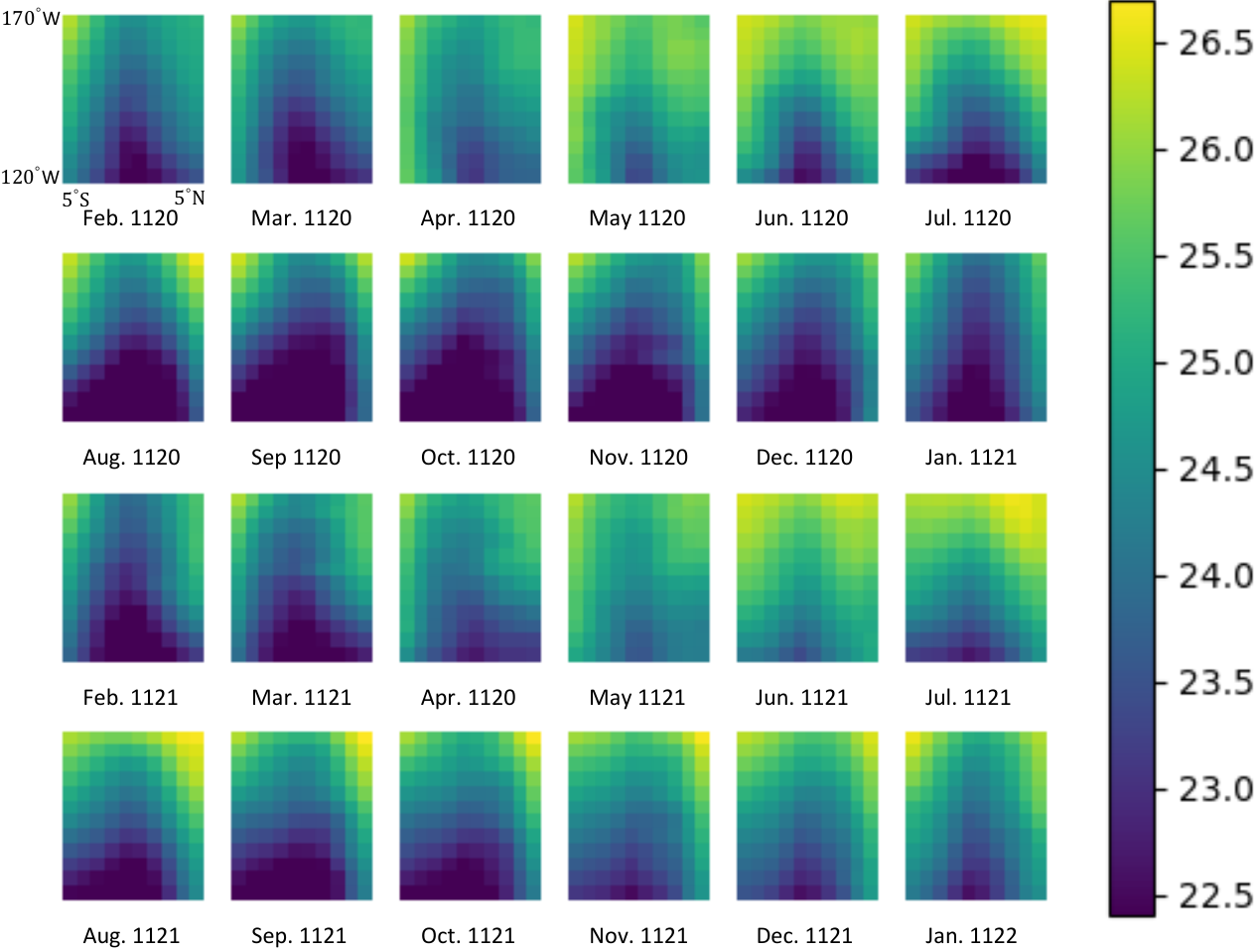}
    \label{fig:CCSM4_truth}
    }
    \subfloat[Prediction]{\includegraphics[width = 0.45\linewidth]{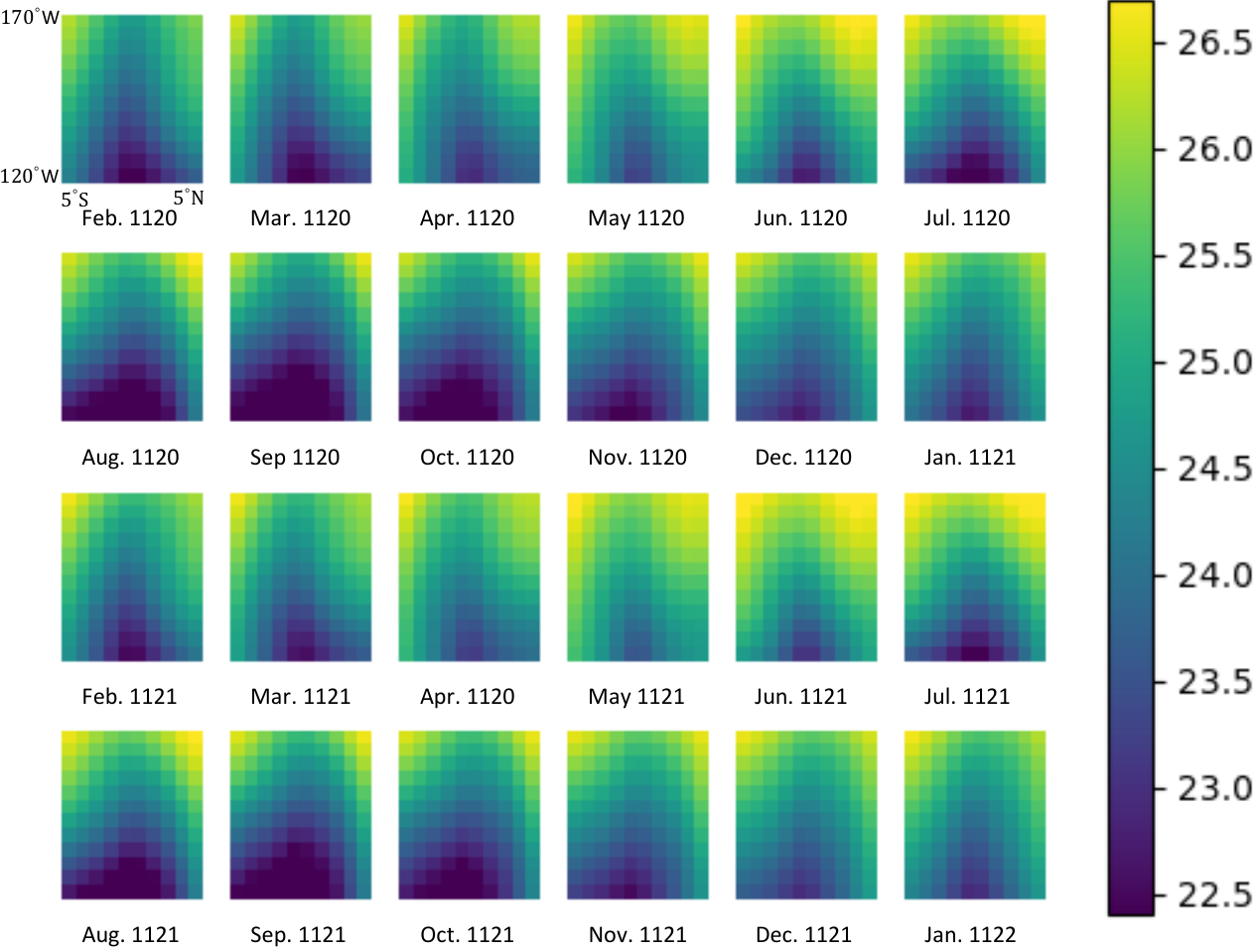}
    \label{fig:CCSM4_pred}}
    \\
    \subfloat[Diff.]{\includegraphics[width = 0.45\linewidth]{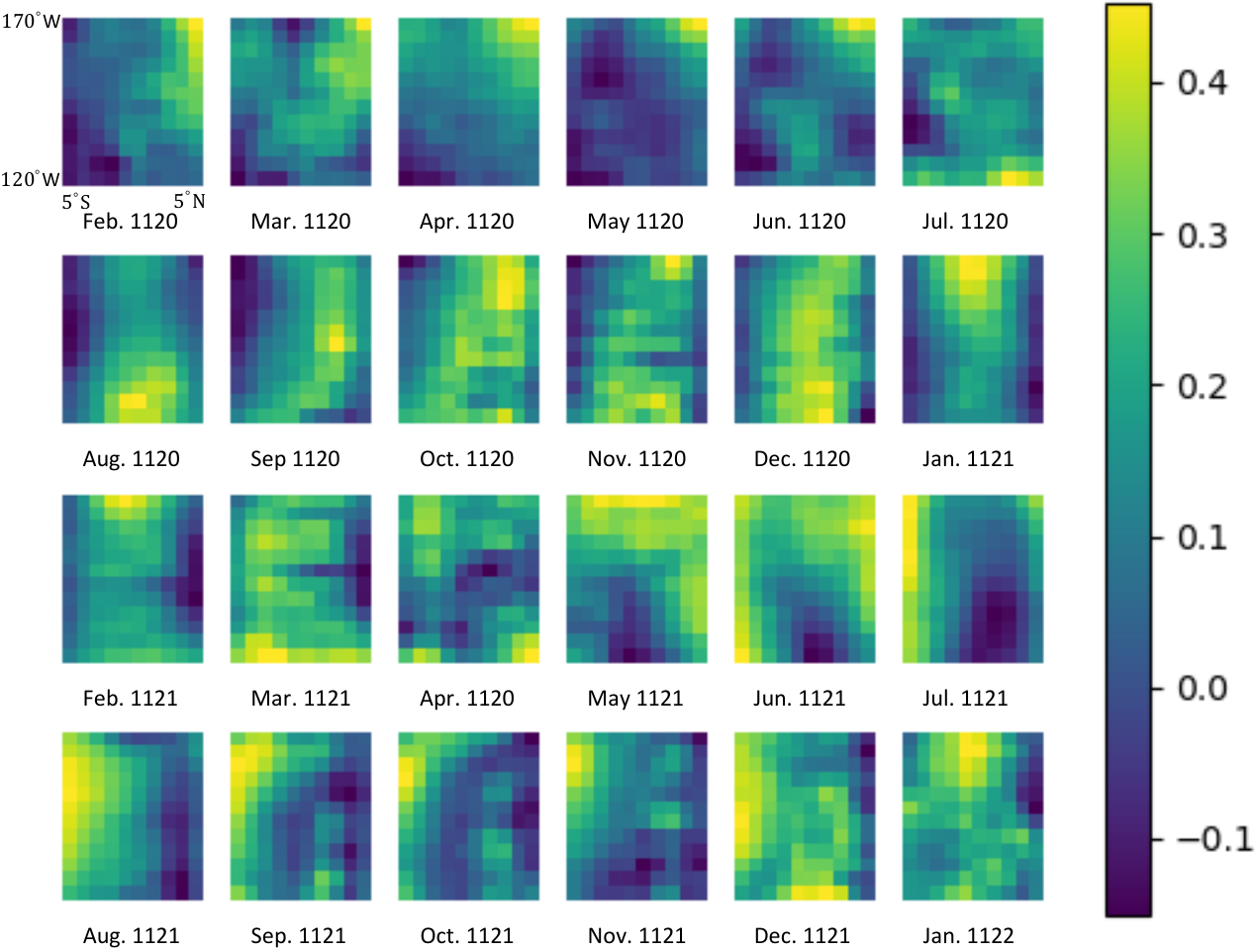}
    \label{fig:CCSM4_diff}}
    \subfloat[RMSE and PC]{\includegraphics[width = 0.45\linewidth]{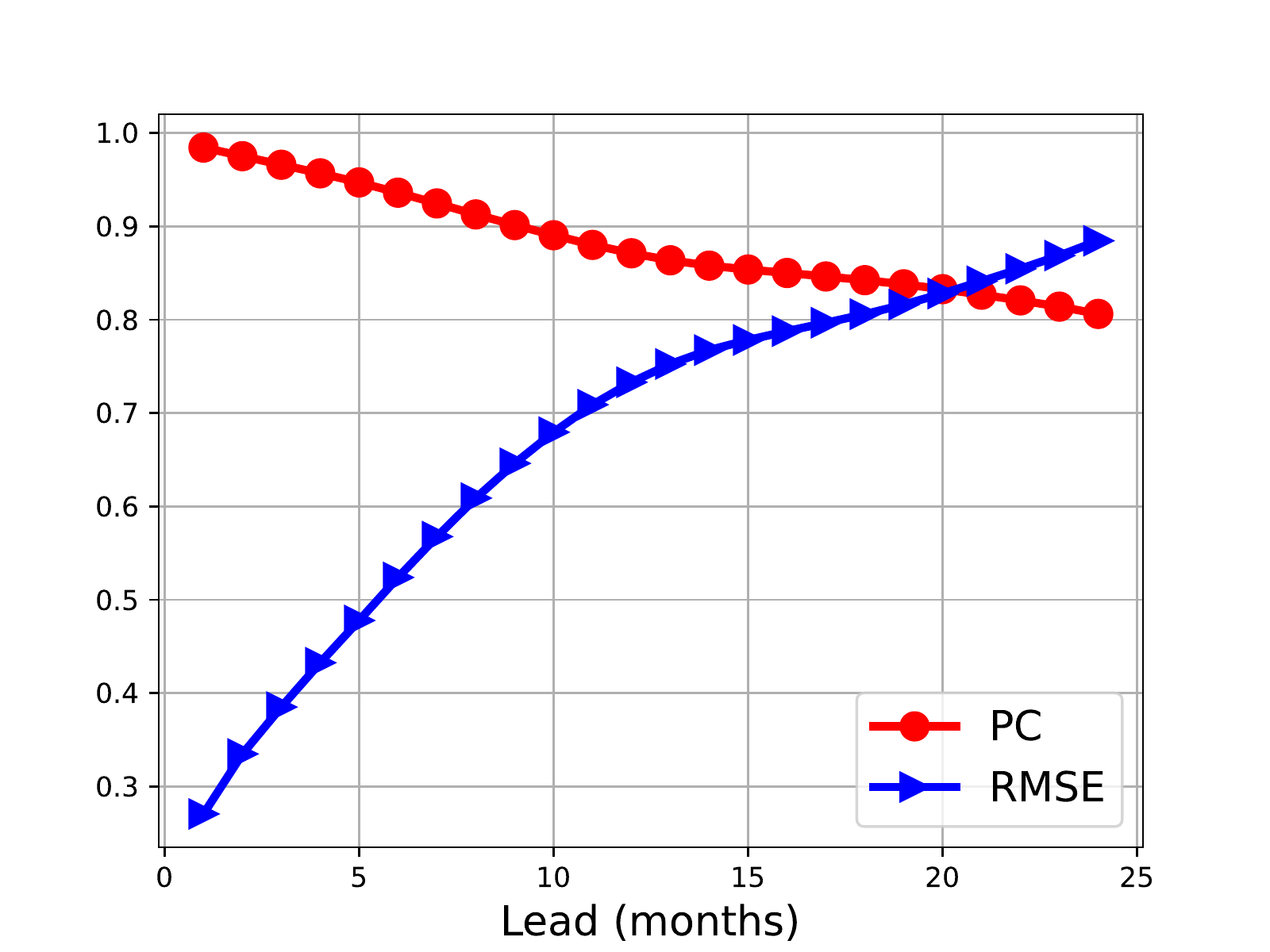}
    \label{fig:CCSM4_PC_RMSE_per_pixel}}
    \caption{Performance on the ENSO region spatio-temporal sequence prediction task. \protect \subref{fig:CCSM4_truth}: Sample ground truth of the ENSO region starting from February 1120. \protect \subref{fig:CCSM4_pred}: Sample prediction of the ENSO region starting from February 1120.
    \protect \subref{fig:CCSM4_diff}: Sample difference between the ground truth and prediction of the ENSO region start from February 1120. \protect \subref{fig:CCSM4_PC_RMSE_per_pixel} RMSE per grid cell and PC as a function of lead time computed in the testing period.}
    \label{fig:CCSM4_1}
\end{figure}

\section*{Results and discussions}

\subsection*{Experimental setup}
Experiment results and discussions of the ConvGRU network on various global climate and atmospheric reanalysis datasets are presented. The reanalysis datasets used consist of two SST datasets and one air temperature dataset. The performance of the ConvGRU network is evaluated on the ENSO region spatio-temporal sequence prediction task for the SST datasets. Additionally, the performance is compared with several existing models on a downstream task of predicting the Ni\~no $3.4$ index, which is calculated based on the aforementioned ENSO region spatio-temporal sequence prediction task. The ConvGRU network is also evaluated on the spatio-temporal sequence prediction task for the air temperature dataset, which covers almost $2/3$ of the global surface. 

For the numerical experiments, the ConvGRU network is implemented using PyTorch~\cite{paszke2019pytorch}. The experiments are conducted on a Linux server equipped with a single GPU, either NVIDIA GeForce GTX 1080Ti or NVIDIA RTX A6000. The implementation codes of the ConvGRU network are publicly available, and more detailed information can be found in the data availability section.

\subsection*{CCSM4 dataset}

\begin{figure}[t!]
    \centering
    \subfloat[PC vs. models using SST maps of the Pacific]{\includegraphics[width = 0.4\linewidth]{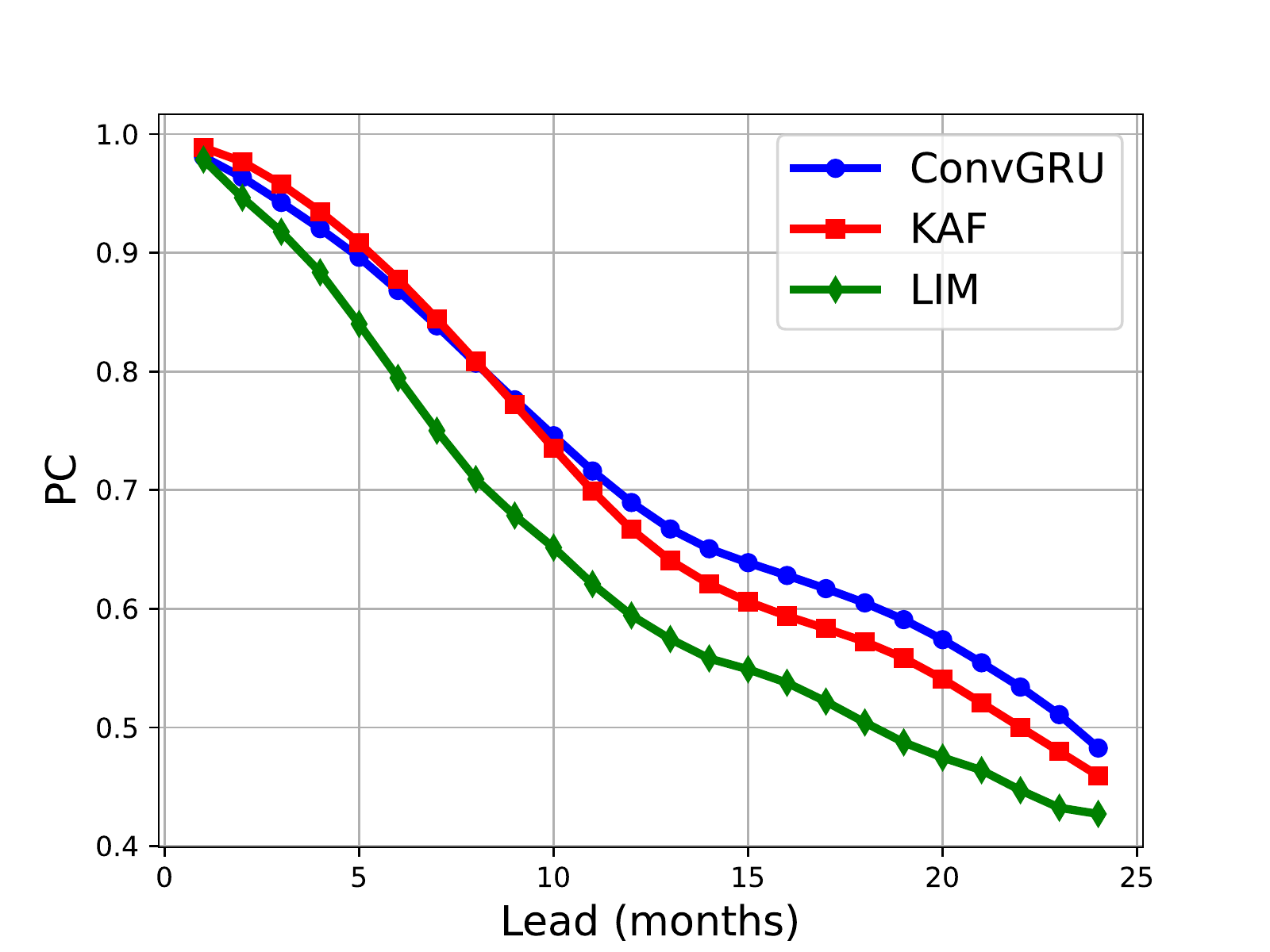}
    \label{fig:CCSM4_PC_gridded}
    }
    \subfloat[PC vs. models using mean SSTs of the ENSO region]{\includegraphics[width = 0.4\linewidth]{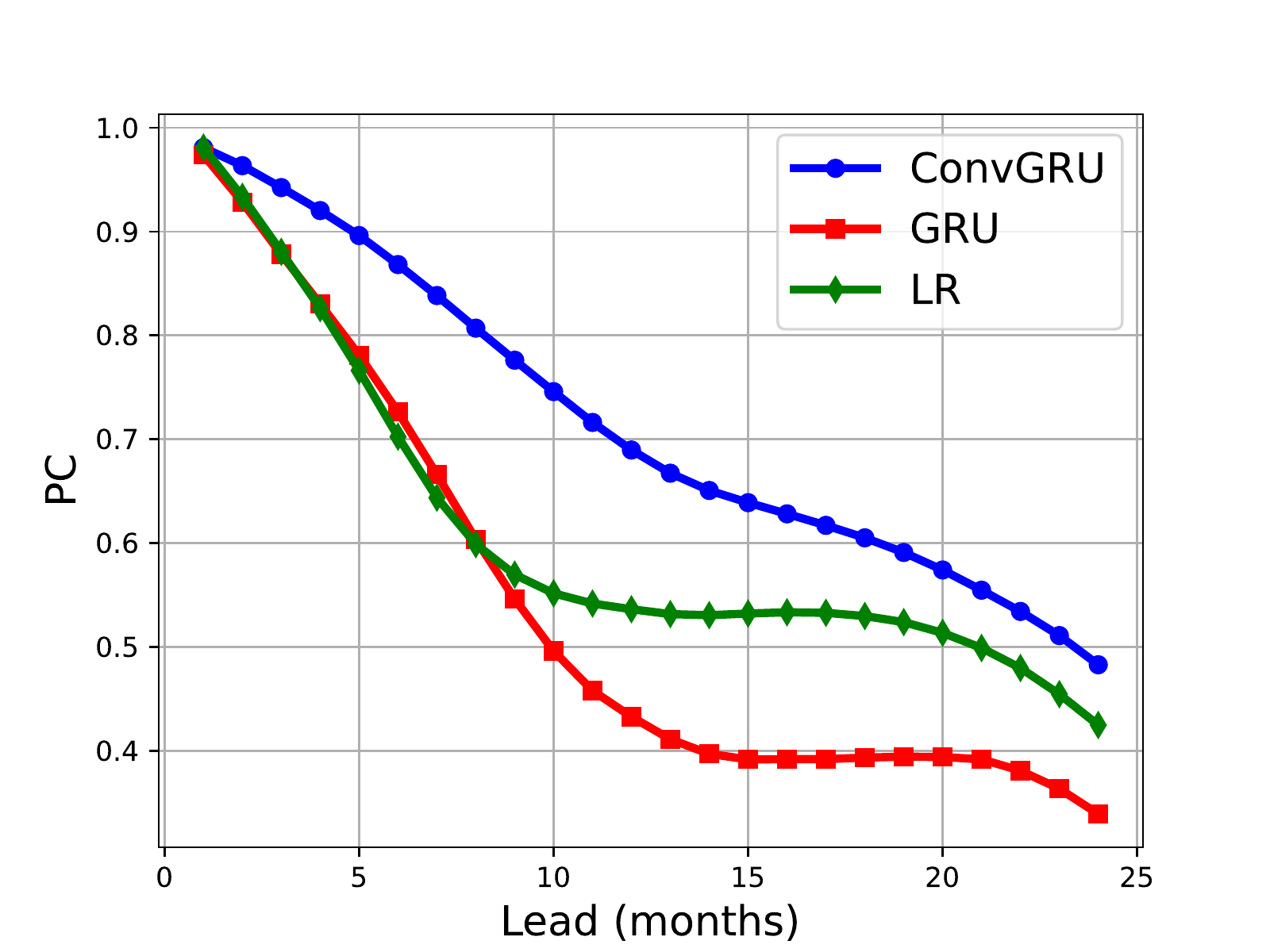}
    \label{fig:CCSM4_PC_1d}}
    \\
    \subfloat[RMSE vs. models using SST maps of the Pacific]{\includegraphics[width = 0.4\linewidth]{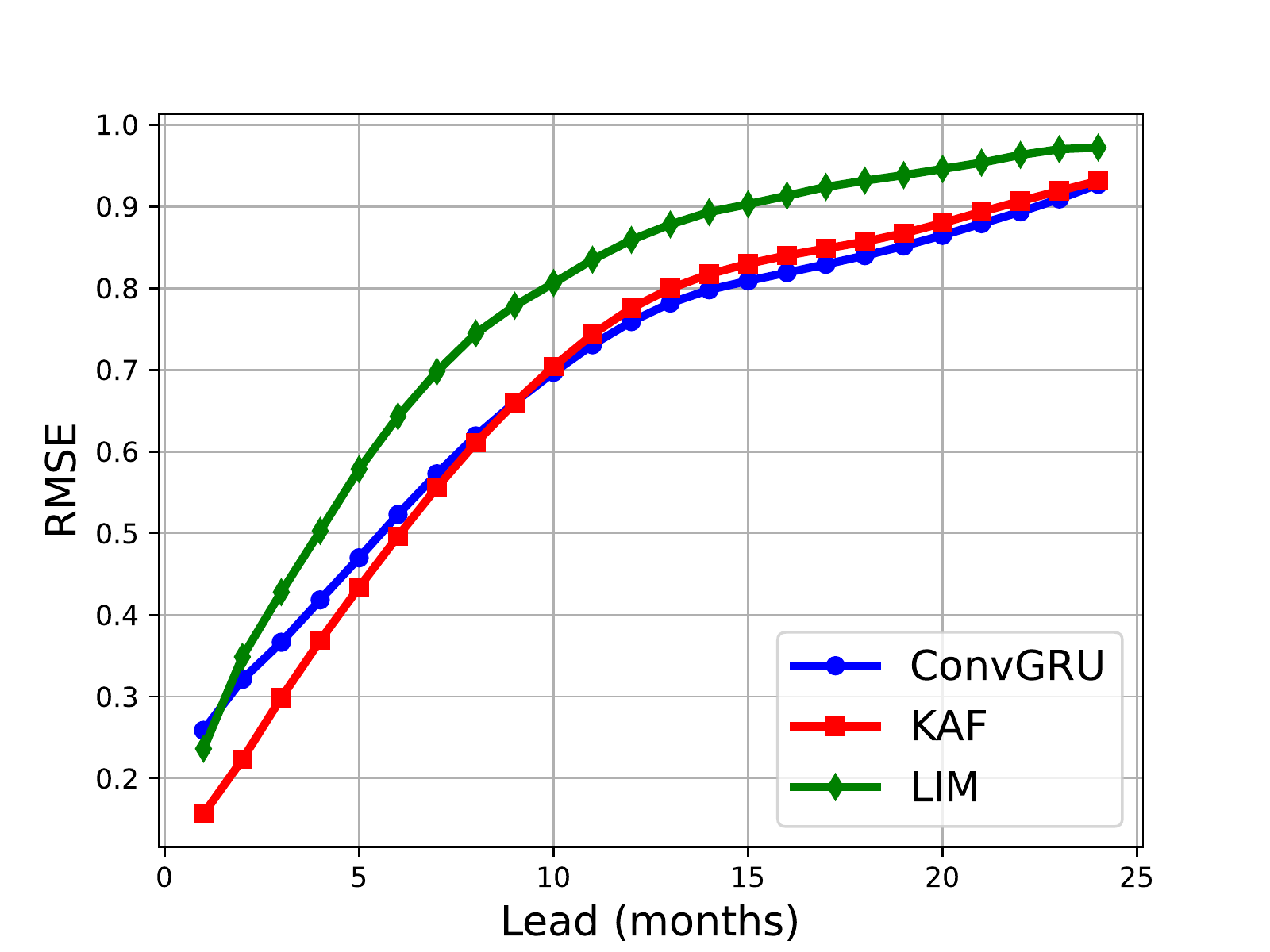}
    \label{fig:CCSM4_RMSE_gridded}}
    \subfloat[RMSE vs. models using mean SSTs of the ENSO region]{\includegraphics[width = 0.4\linewidth]{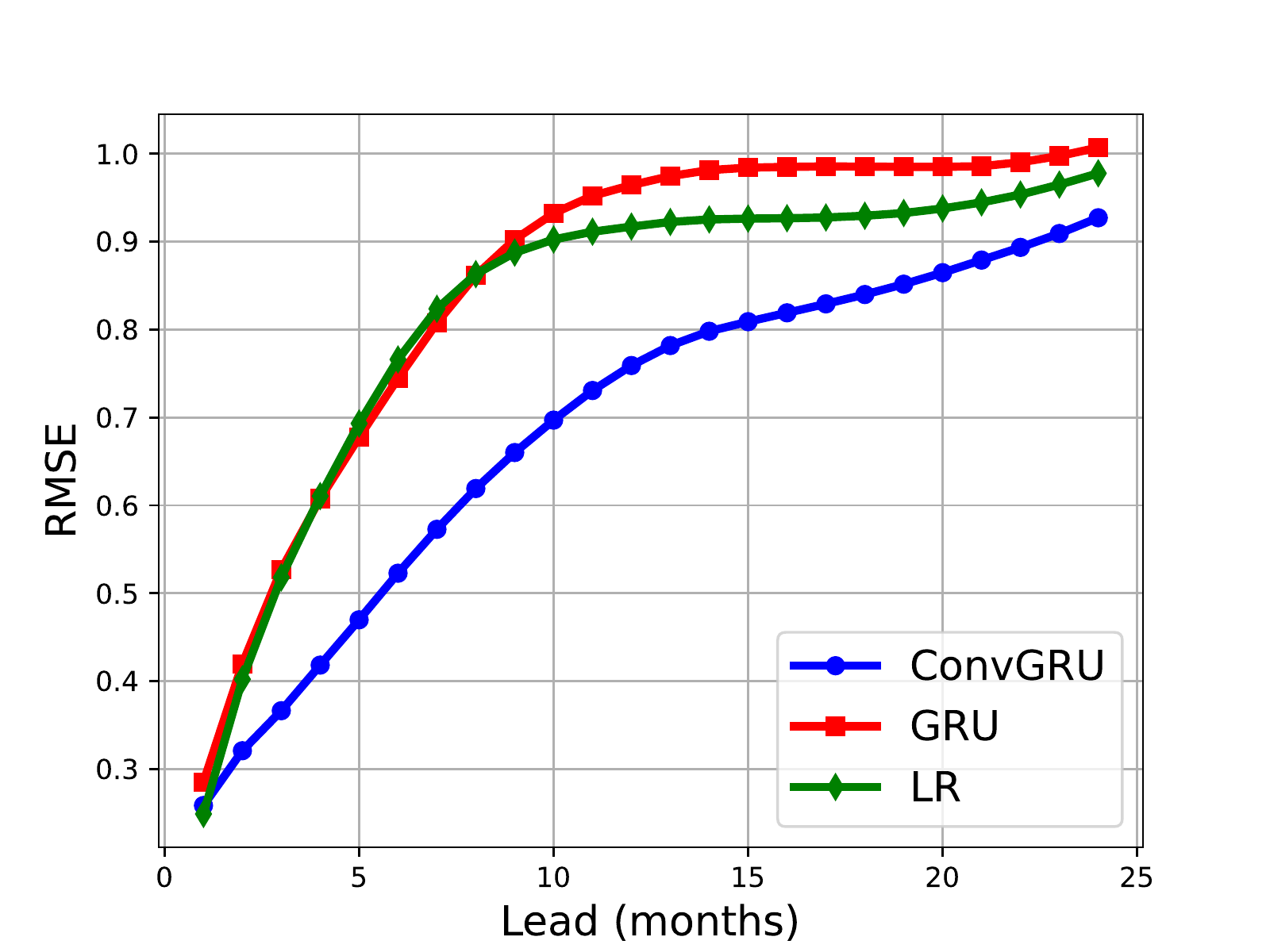}
    \label{fig:CCSM4_RMSE_1d}}
    \caption{Performance of the ConvGRU network against other models on predicting the Ni\~no $3.4$ index in the CCSM4 dataset during 1100-1300. \protect \subref{fig:CCSM4_PC_gridded} and \protect \subref{fig:CCSM4_RMSE_gridded} PC and RMSE, respectively, as a function of lead time, compared to KAF and LIM. \protect \subref{fig:CCSM4_PC_1d} and \protect \subref{fig:CCSM4_RMSE_1d} PC and RMSE, respectively, as a function of lead time, compared to Seq2Seq with GRU and LR.}
    \label{fig:CCSM4_2}
\end{figure}

The CCSM4 dataset is a modeled SST dataset derived from a $1300$-year, pre-industrial control integration of the Community Climate System Model version~$4$ (CCSM4)~\cite{gent2011community}. The dataset is sampled and averaged monthly on the model's native ocean grid with a normal resolution of approximately $1^\circ\times 0.5^\circ$ (longitude-latitude). The SST maps of the Pacific and ENSO regions are extracted from specific longitude-latitude boxes. The Pacific region covers $16^\circ \text{E}$-$56^\circ \text{W}$ and $69^\circ \text{S}$-$32^\circ \text{N}$ ($256 \times 256$ grid), while the ENSO region covers $170^\circ$-$120^\circ \text{W}$ and $5^\circ \text{S}$-$5^\circ \text{N}$ ($45 \times 38$ grid). To reduce computational complexity and GPU memory usage, the SST maps of the Pacific and ENSO regions are down-sampled to $64 \times 64$ and $12 \times 10$ grids, respectively. The CCSM4 dataset is split into disjoint training and testing data periods, with years 1-1099 allocated for training and years 1100-1300 for testing. 

For experiments on the CCSM4 dataset, a 3-layer ConvGRU network is implemented with parameters provided in the \verb|NetParams.py| in the \verb|ConvGRU_CCSM4| folder. The condition range ($J$) and the prediction range ($K$) are set to 48 and 24 months, respectively.

Figure~\ref{fig:CCSM4_1} illustrates the performance of the ConvGRU network on the ENSO region spatio-temporal sequence prediction task. Figures~\ref{fig:CCSM4_truth}, \ref{fig:CCSM4_pred} and \ref{fig:CCSM4_diff} include a sample comparison, starting from February 1120, between the ground truth and the network's prediction for the ENSO region. The patterns in both the ground truth and the prediction exhibit high similarity.  Figure~\ref{fig:CCSM4_PC_RMSE_per_pixel} presents the prediction skill assessed using RMSE and PC metrics over the entire testing period as a function of lead time. The RMSE values are averaged over all possible start points in the testing data split and grid cells of the ENSO region. For PC, the ground truths and predictions are vectorized and concatenated over all possible start points in the testing data split. The RMSE and PC results demonstrate the high correlation and low error characteristics of predictions generated by the ConvGRU network.

Next, the performance of the ConvGRU network is compared with existing models for predicting the Ni\~no $3.4$ index in the CCSM4 dataset. Here Ni\~no $3.4$ indices mean SST anomalies relative to monthly climatology (average SST) of the ENSO region. The models selected for comparison include KAF~\cite{wang2020extended, zhao2016analog, burov2021kernel}, LIM~\cite{wang2020extended, penland1993prediction, penland1995optimal}, Seq2Seq with GRU~\cite{cho2014properties, chung2014empirical}, and LR. KAF and LIM utilize SST maps of the Pacific region as input (predictor) variables, while Seq2Seq with GRU and LR use mean SSTs of the ENSO region. Seq2Seq with GRU is implemented using the DeepAR model~\cite{salinas2020deepar} from the GluonTS package~\cite{alexandrov2020gluonts}, with a 1-layer GRU network with a 20-dimensional hidden state, and the condition and prediction ranges ($J$ and $K$) the same as the ConvGRU network. For LR, $K=24$ separate models are trained for lead months 1-24, using $J=48$ months lagged mean SSTs of the ENSO region (including the current month) as input features.

Figure~\ref{fig:CCSM4_2} illustrates the performance of the ConvGRU network compared against other models in predicting the Ni\~no $3.4$ index in the CCSM4 dataset over the testing period of 1100-1300, with the training period from 1 to 1099. A threshold of PC~$=0.6$ is commonly used to differentiate useful from non-useful predictions~\cite{wang2020extended}. The comparison demonstrates that although the performance of the ConvGRU network deteriorates with longer lead times, it consistently outperforms the competing models in terms of both PC and RMSE, particularly in the long-term prediction range. When considering the useful prediction range using the PC threshold of $0.6$, the ConvGRU network achieves the longest useful range of 18-19 months, surpassing KAF, LIM, Seq2Seq with GRU, and LR by 3-4, 6-7, 10-11, and 10-11 months, respectively. 

\subsection*{NOAA-GDFL-SPEAR dataset}

\begin{figure}[t!]
    \centering
    \subfloat[PC]{\includegraphics[width = 0.4\linewidth]{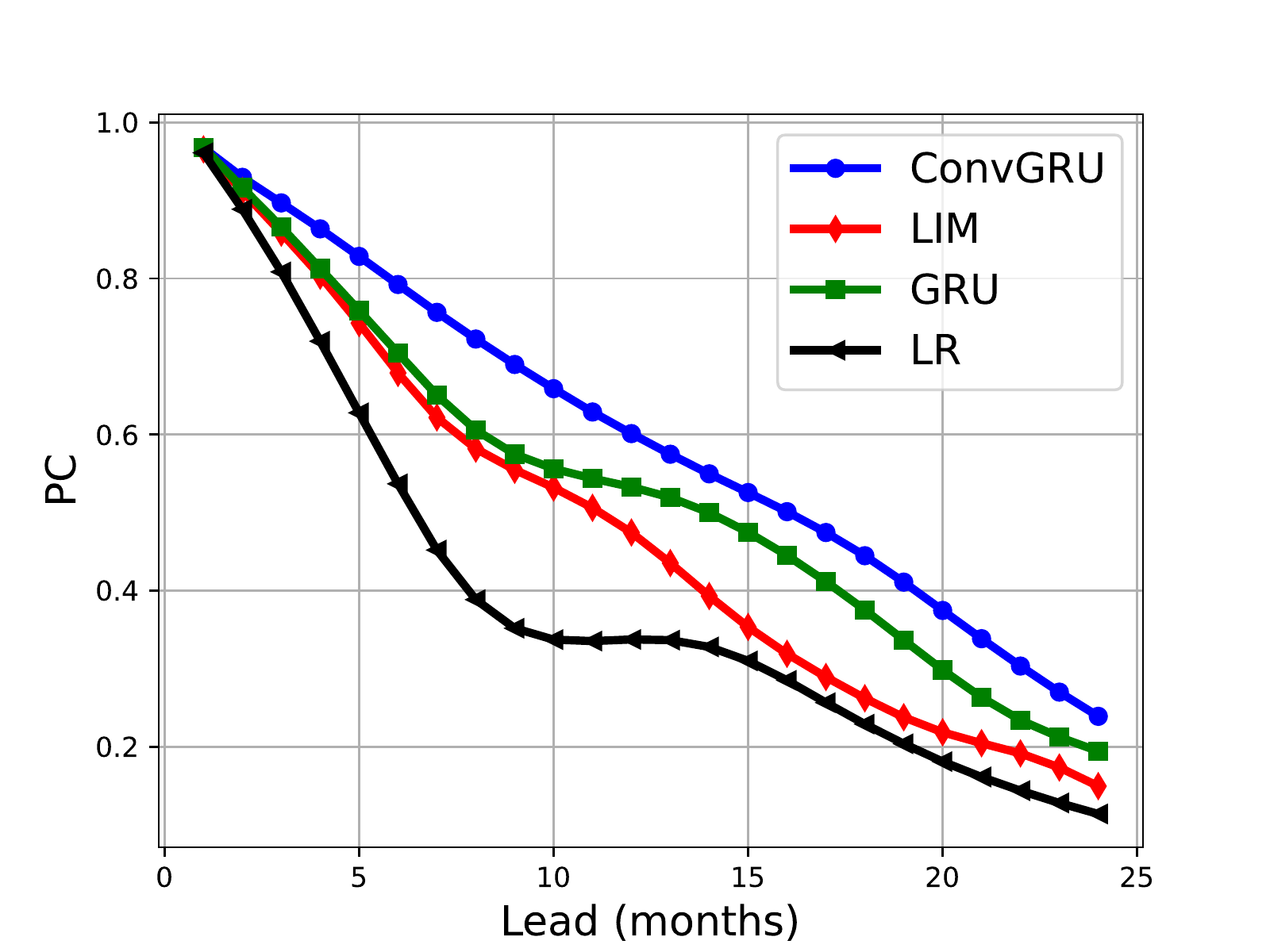}
    \label{fig:SPEAR_PC}
    }
    \subfloat[RMSE]{\includegraphics[width = 0.4\linewidth]{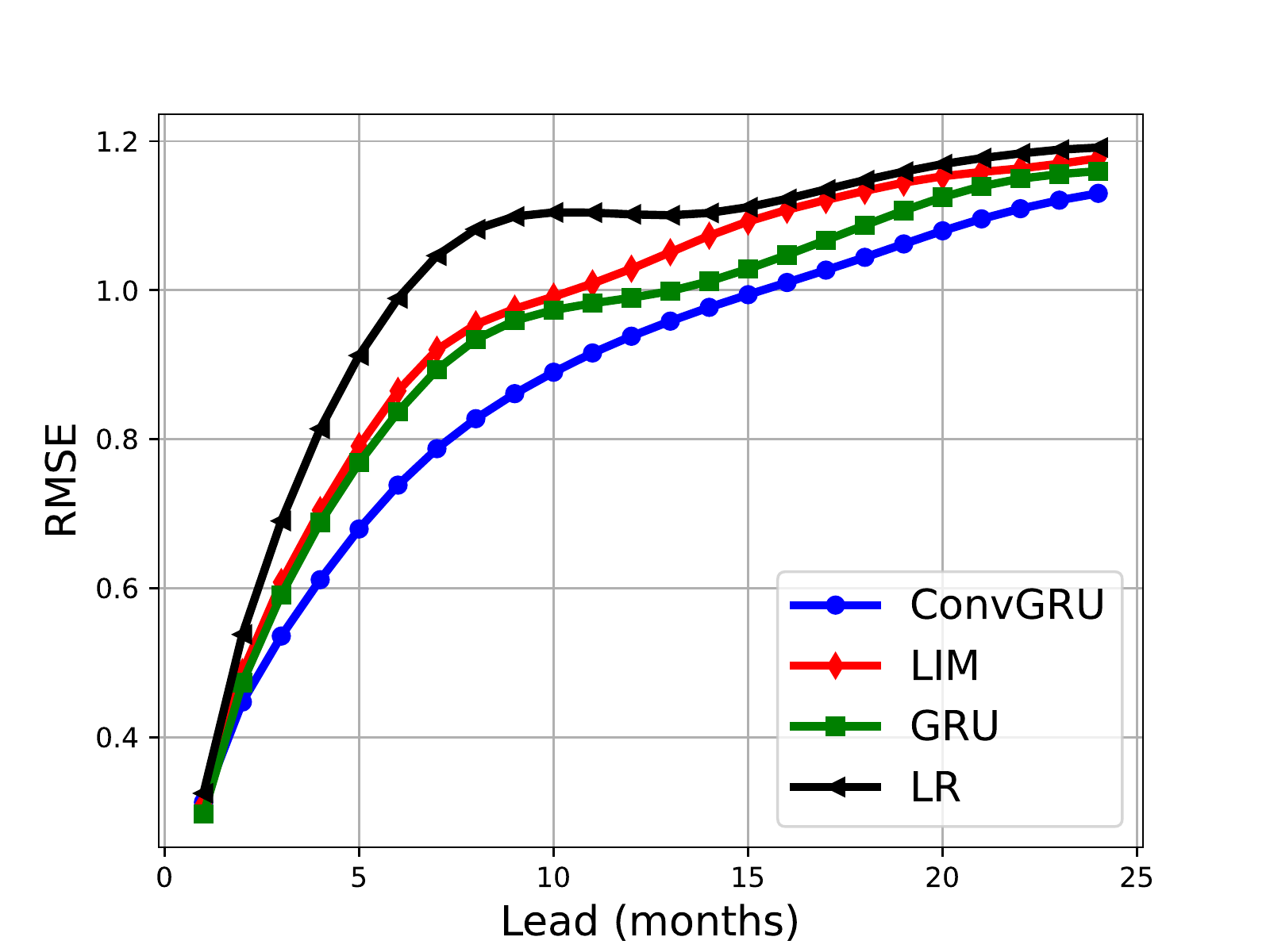}
    \label{fig:SPEAR_RMSE}}
    \caption{Performance averaged over 30 ensembles of the ConvGRU network against other models on predicting the Ni\~no 3.4 index in the NOAA-GDFL-SPEAR dataset during 2051-2100. \protect \subref{fig:SPEAR_PC} and \protect \subref{fig:SPEAR_RMSE} PC and RMSE, respectively, as a function of lead time, compared to LIM, Seq2Seq with GRU, and LR.}
    \label{fig:SPEAR}
\end{figure}

\begin{figure}[htb!]
    \centering
    \subfloat[Ground truth]{\includegraphics[width = 0.4\linewidth]{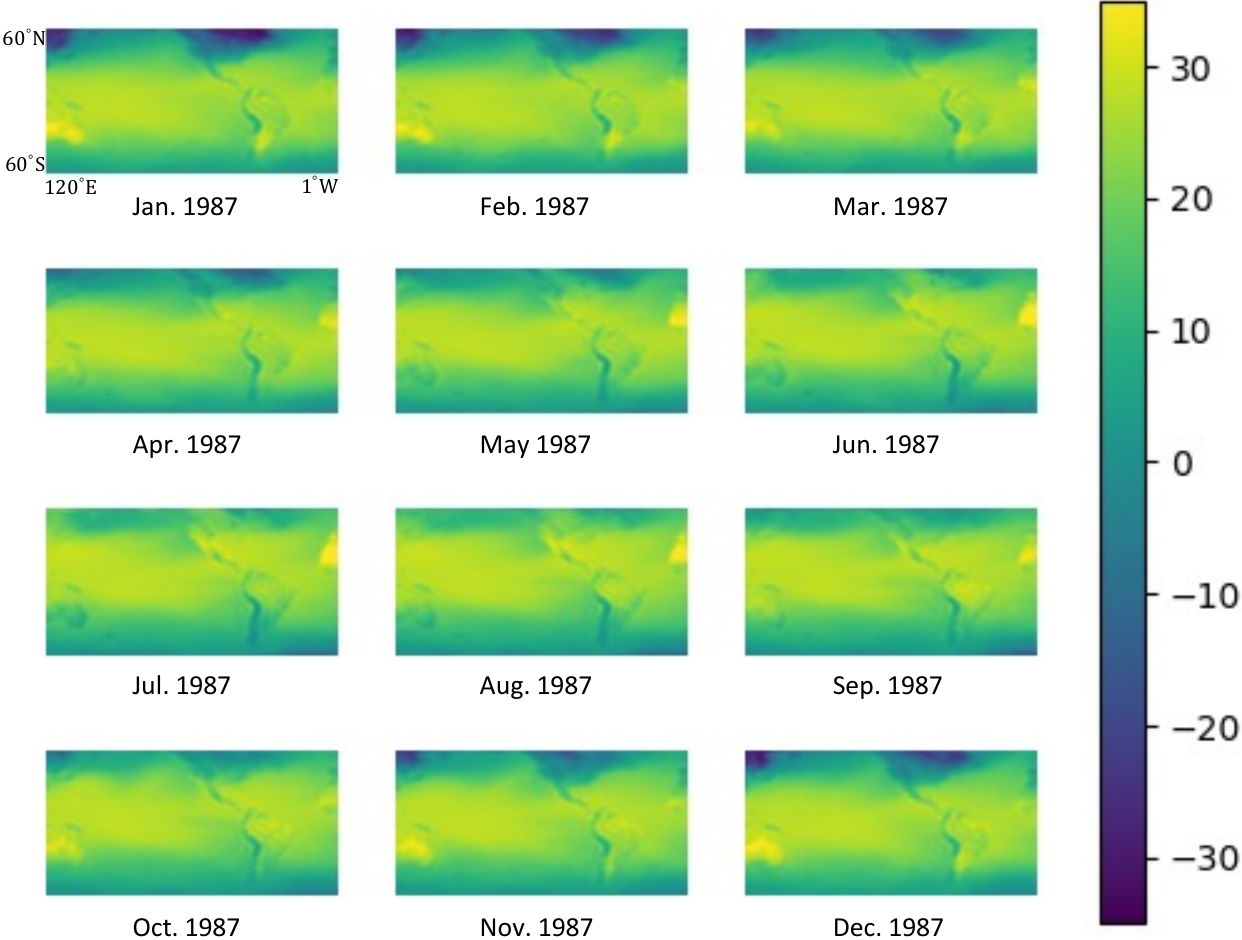}
    \label{fig:Air_Temp_truth}
    }
    \subfloat[Prediction]{\includegraphics[width = 0.4\linewidth]{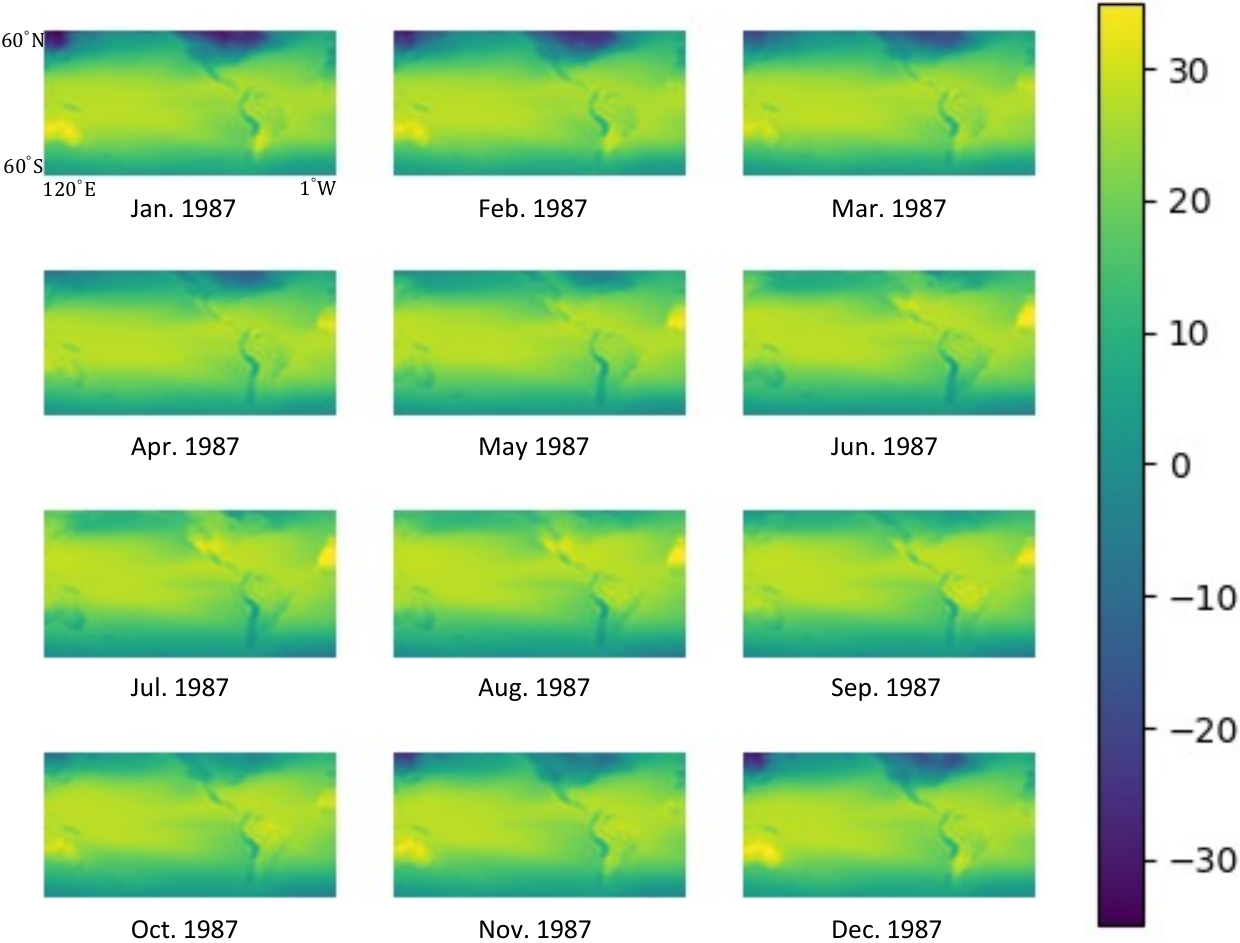}
    \label{fig:Air_Temp_forecast}}\\
    \subfloat[Diff.]{\includegraphics[width = 0.4\linewidth]{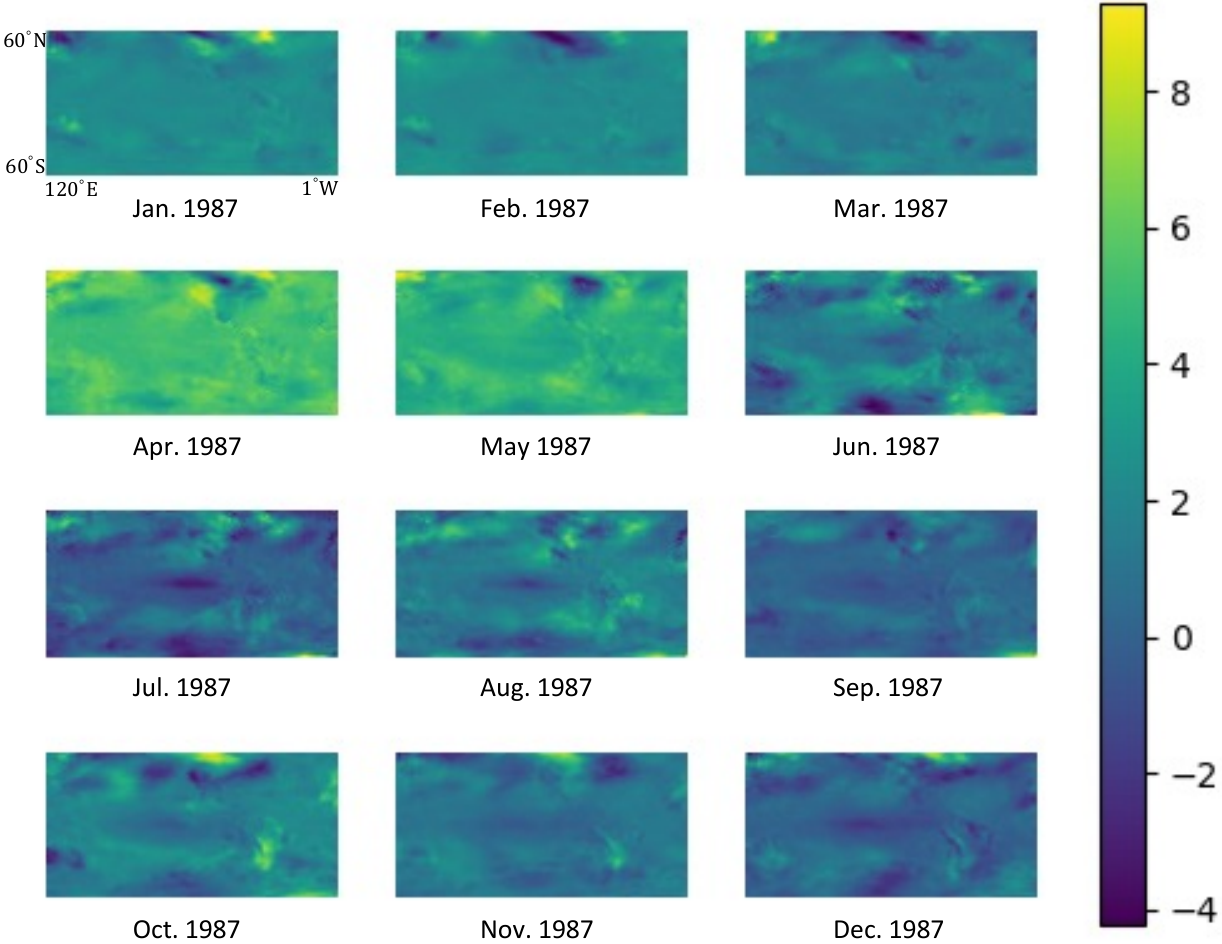}
    \label{fig:Air_Temp_diff}}\\
    \subfloat[PC]{\includegraphics[width = 0.365\linewidth]{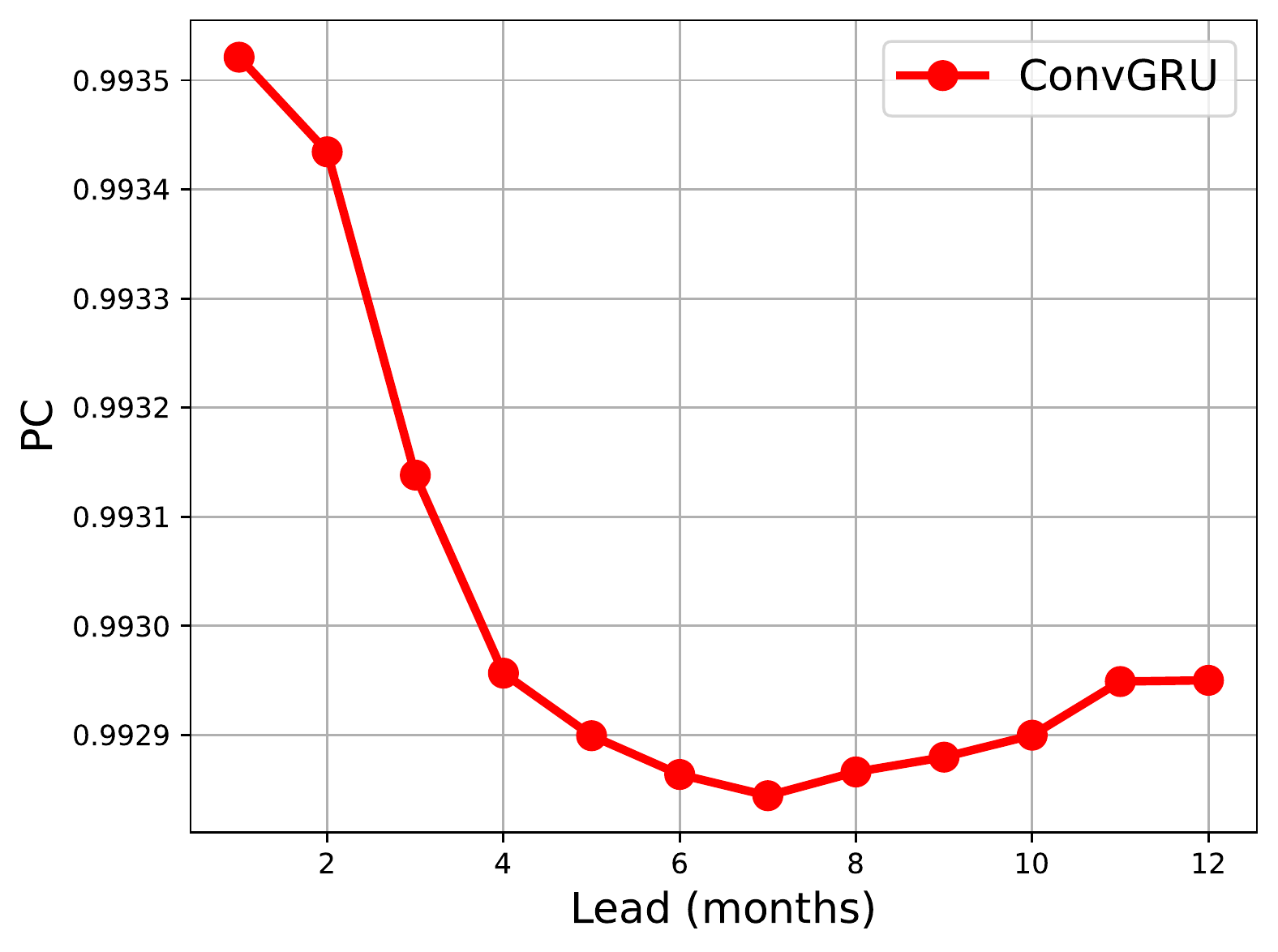}
    \label{fig:Air_Temp_PC_per_pixel}}
    \subfloat[RMSE]{\includegraphics[width = 0.4\linewidth]{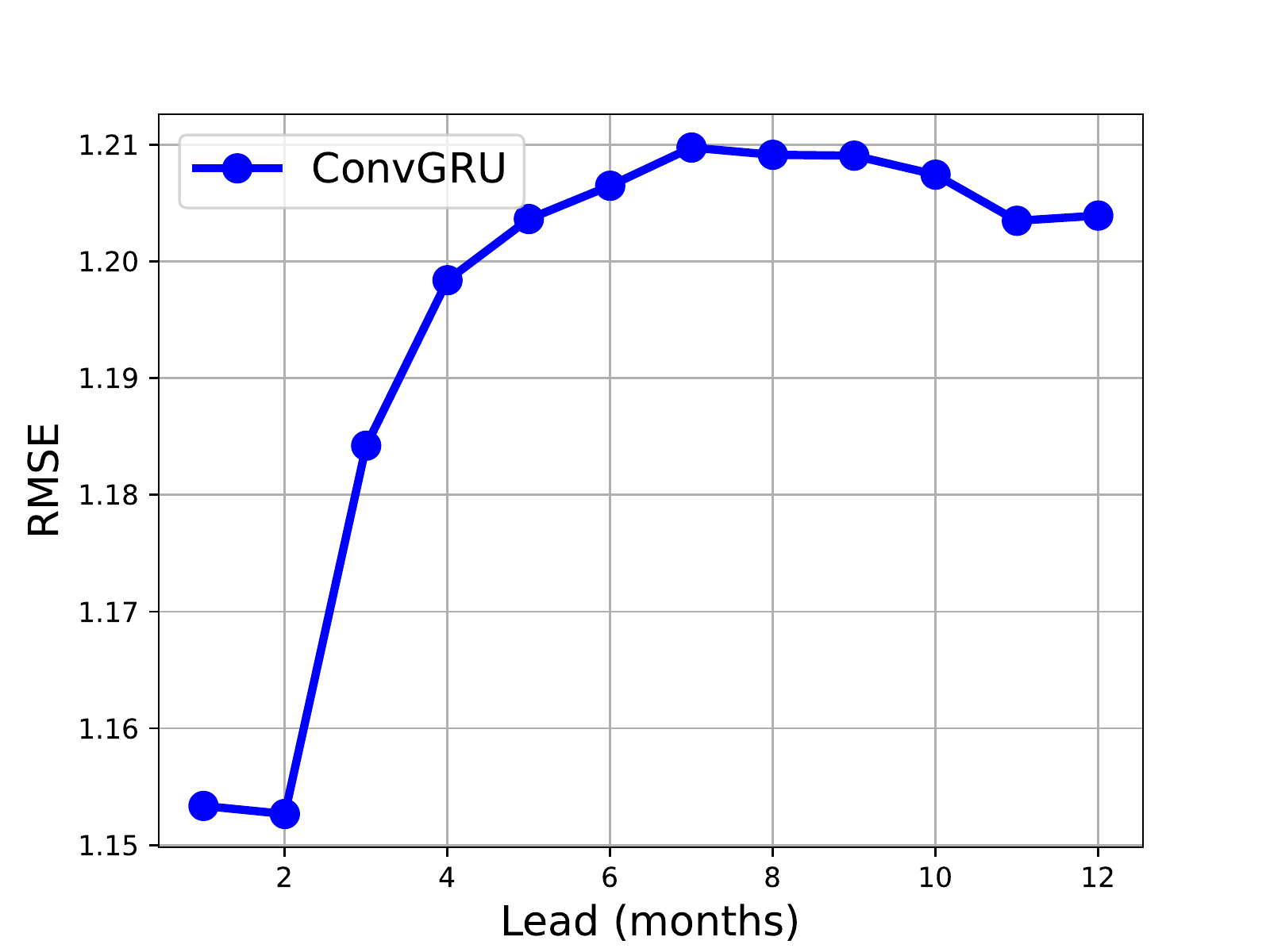}
    \label{fig:Air_Temp_RMSE_per_pixel}}
    \caption{Performance on the air temperature spatio-temporal sequence prediction task. \protect \subref{fig:Air_Temp_truth} Sample ground truth starting from January 1987. \protect \subref{fig:Air_Temp_forecast} Sample prediction starting from January 1987.
    \protect \subref{fig:Air_Temp_diff} Sample difference between ground truth and prediction starting from January 1987.
    \protect \subref{fig:Air_Temp_PC_per_pixel} PC as a function of lead time computed in the testing period. \protect \subref{fig:Air_Temp_RMSE_per_pixel} RMSE per grid cell as a function of lead time computed in the testing period.}
    \label{fig:Air_Temp}
\end{figure}

The NOAA-GDFL-SPEAR dataset used in the numerical experiment is a simulated monthly averaged SST dataset with a nominal resolution of  $1^\circ \times 1^\circ$ (longitude-latitude) from the GFDL SPEAR large ensembles. This includes 30-member ensembles of climate change simulations covering the period 1921-2100 using the SPEAR-MED climate model~\cite{delworth2020spear}. The simulations are forced with historical radiative forcings from 1921 to 2014 and SSP5-8.5 projected radiative forcings~\cite{van2014new,o2016scenario} from 2015 to 2100. The SST maps of the Pacific and ENSO region are extracted from the longitude-latitude boxes $150^\circ \text{E}$-$82^\circ \text{W}$, $69^\circ \text{S}$-$59^\circ \text{N}$ ($128 \times 128$ grid) and $170^\circ$-$120^\circ \text{W}$, $5^\circ \text{S}$-$5^\circ \text{N}$ ($50 \times 10$ grid), respectively. Similar to the previous comparison, the SST maps in both regions are down-sampled to $64 \times 64$ and $25 \times 5$ grids, reducing computational complexity and GPU memory usage. 

The NOAA-GDFL-SPEAR dataset in each ensemble is divided into disjoint
training and testing data splits, with the training period covering years 1921-2050 and the testing period covering years 2051-2100. For experiments on this dataset, a 3-layer ConvGRU network is implemented using the parameters specified in the \verb|NetParams.py| file in the \verb|ConvGRU_SPEAR| folder. The ConvGRU network is trained using data from all 30 ensembles that end on or before the year 2050, and then tested on data from all 30 ensembles starting from the year 2051. The condition range ($J$) and the prediction range ($K$) are 48 and 24 months, respectively.

Similar to the previous comparison, the performance of the ConvGRU network is compared to several existing models for predicting the Ni\~no $3.4$ index in the NOAA-GDFL-SPEAR dataset. The selected models for comparison include LIM, Seq2Seq with GRU, and LR. For LIM and LR, separate models are trained for each ensemble in the dataset, since the NOAA-GDFL-SPEAR dataset contains data from 30 ensembles, and the metrics computed in the testing period are averaged over all ensembles. Seq2Seq with GRU utilizes the DeepAR model to handle multiple time series with a single model, and the detailed and fine-tuned settings remain the same in the CCSM4 dataset. 

Figure~\ref{fig:SPEAR} presents the performance averaged over 30 ensembles of the ConvGRU network compared against other models in predicting the Ni\~no $3.4$ index in the NOAA-GDFL-SPEAR dataset over the testing period of 2051-2100, using the training period of 1921-2051. The results of the experiments demonstrate that the ConvGRU network significantly outperforms the competing models in terms of both PC and RMSE, with the longest useful range of 12 months, surpassing LIM, Seq2Seq with GRU, and LR by 4-5, 4, and 7 months, respectively.

\subsection*{NOAA-CIRES air temperature dataset}

The NOAA-CIRES air temperature dataset used in this experiment is a simulated monthly ensemble mean air temperature dataset at the 2m level with a nominal resolution of approximately $2^\circ\times 2^\circ$ (longitude-latitude). It is from the NOAA-CIRES 20th-Century Reanalysis Version 2c~\cite{compo2011twentieth,krueger2013inconsistencies}, which provides comprehensive global atmospheric circulation dataset spanning the years 1850-2014. For the NOAA-CIRES air temperature dataset, we aim to demonstrate that the ConvGRU network is capable of accurately predicting other climate and atmospheric spatio-temporal sequence beyond the ENSO region. For this experiment, the target region is the longitude-latitude box $120^\circ \text{E}-1^\circ \text{W}$, $60^\circ \text{N}-60^\circ \text{S}$ ($128\times 64$ grid), covering almost two-thirds of the total global surface. 

The NOAA-CIRES air temperature dataset is divided into disjoint training and testing periods, with the training period covering the years 1851-1980 and testing period covering 1981-2014. In this experiment, a 3-layer ConvGRU network is implemented using the parameters specified in the \verb|NetParams.py| file in the \verb|AirTmp_M| folder. The condition range ($J$) and the prediction range ($K$) are set to 24 and 12 months, respectively.

Figure~\ref{fig:Air_Temp} illustrates the performance of the ConvGRU network on the air temperature spatio-temporal sequence prediction task. Figures~\ref{fig:Air_Temp_truth},~\ref{fig:Air_Temp_forecast}, and~\ref{fig:Air_Temp_diff} presents a sample comparison starting from January 1987 between the ground truth and the prediction, revealing a very similar pattern in both the ground truth and the prediction. Figures~\ref{fig:Air_Temp_PC_per_pixel} and \ref{fig:Air_Temp_RMSE_per_pixel} present the prediction skill assessed using RMSE and PC, similar to Fig.~\ref{fig:CCSM4_1}, over the entire testing data split and as a function of lead time. The PC result demonstrates a significantly high correlation of over $99\%$ between the ground truth and the prediction within a prediction range of 12 months. The RMSE ranges from approximately $1.1$ \textcelsius~to $1.2$ \textcelsius.

\section*{Conclusion}

We addressed the ENSO region spatio-temporal sequence prediction problem by proposing a modified ConvGRU network, as well as its downstream task of predicting the Ni\~no $3.4$ index. The ConvGRU network incorporated 2-D convolutional layers within a ConvGRU cell and employed an encoder-decoder Seq2Seq structure, offering advantages over existing models such as LR, LIMs, KAF, and Seq2Seq with GRU. These advantages include the ability to output future SST maps of the ENSO region, rather than ENSO indices, and modelling approximate nonlinear dynamics. 

Through experiments on various climate and atmospheric reanalysis datasets, we demonstrated the effectiveness of the ConvGRU network in predicting future SST maps in the ENSO region. The ConvGRU network outperformed existing models in various scenarios, including the Ni\~no $3.4$ index prediction, showcasing its capabilities in downstream applications. We also evaluated the performance of the network in predicting other climate-related tasks, such as predicting monthly air temperature over a large portion of the global surface, which further demonstrate its potential for accurate spatio-temporal sequence predictions.

Overall, the proposed ConvGRU network offers a promising approach for ENSO region spatio-temporal sequence prediction and hold potential for advancing climate and atmospheric prediction and related domains. There are opportunities for further exploration, such as variants or successors of the ConvGRU network and leveraging Transformer-based models to enhance prediction skills. Additionally, this study lays the foundation for constructing real-time forecast systems with monthly updates based on the availability of observational data.

\section*{Data availability}
The datasets generated and analyzed during this study are available from the corresponding author on reasonable request. Additionally, the codes and detailed information about the datasets can be found at the following public GitHub repository: \url{https://github.com/LingdaWang/ConvGRU_ENSO_Forecast}.

\bibliography{refs}

\section*{Acknowledgements}
This study was partially based on the research supported by the Discovery Partners Institute (DPI) Science Team Seed Grant Program; DPI is part of the University of Illinois System. It was partially supported by NSF OAC-1934757, the NSF Graduate Research Fellowship Program (DGE 21-46756), the Alfred P. Sloan Foundation, and the Sloan University Center of Exemplary Mentoring at Illinois. The authors extend their gratitude to Prof. Dimitrios Giannakis at Dartmouth College for providing the preprocessed CCSM4 dataset and the results of KAF.

\section*{Author contributions}
Z.Z., R.S., and V.H. designed the research. L.W., Z.Z., R.S., and V.H. proposed the methodology, L.W. and S.A. performed numerical experiments. L.W. wrote the main manuscript. All authors reviewed and edited the manuscript.

\section*{Competing interests}
The authors declare no competing interests.

\section*{Additional information}

\textbf{Correspondence} and requests for materials should be addressed to L.W., R.S., or Z.Z.

\end{document}